\documentclass{iopart}
\usepackage{iopams}
\usepackage{amssymb}
\usepackage{bm}
\usepackage{dsfont}
\usepackage{graphicx}
\usepackage{subfigure}
\usepackage{dcolumn}
\usepackage{braket}
\usepackage[usenames,dvipsnames,svgnames,table]{xcolor}
\usepackage{cite} 
\definecolor{myred}{RGB}{212,0,0} 
\definecolor{myblue}{RGB}{0,0,128}
\definecolor{mygreen}{RGB}{16,96,16}
\definecolor{vured}{RGB}{137,28,46}
\definecolor{degold}{RGB}{255,204,0}
\definecolor{magic}{RGB}{139,0,139}

\newcommand{\vured}{\color{vured}}
\newcommand{\vublack}{\color{black}}

\newcommand{\rd}{\mathrm{d}}
\newcommand{\bo}{\hat{b}^{\phantom\dag}}
\newcommand{\ba}{\hat{b}^{\dag}}
\newcommand{\ao}{\hat{a}^{\phantom\dag}}
\renewcommand{\aa}{\hat{a}^{\dag}}

\newcommand{\no}{\hat{n}}
\newcommand{\Ho}{\hat{H}}
\newcommand{\Ao}{\hat{A}}
\newcommand{\Bo}{\hat{B}}
\newcommand{\Co}{\hat{C}}
\newcommand{\Vo}{\hat{V}}

\newcommand{\Uo}{\hat{U}}

\newcommand{\Qo}{\hat{Q}}

\newcommand{\la}{\langle}
\newcommand{\ra}{\rangle}
\newcommand{\lla}{\langle\!\langle}
\newcommand{\rra}{\rangle\!\rangle}
\newcommand{\be}{\begin{equation}}
\newcommand{\ee}{\end{equation}}
\newcommand{\bes}{\begin{eqnarray}}
\newcommand{\ees}{\end{eqnarray}}

\renewcommand{\br}{{\bm r}}

\newcommand{\bF}{{\bm F}}

\newcommand{\bk}{{\bm k}}

\newcommand{\bJ}{\mathcal{J}}

\newcommand{\substack}[1]{\mbox{\scriptsize$\!\!\begin{array}{c} #1\end{array}\!\!$}}
\newcommand{\text}[1]{\mathrm{#1}}
\newcommand{\T}{\textstyle}

\begin{document}

\title[High-frequency approximation for periodically driven quantum systems]{High-frequency approximation for periodically driven quantum systems from a Floquet-space perspective}

\author{Andr\'e Eckardt}
\address{Max-Planck-Institut f\"ur Physik komplexer Systeme, N\"othnitzer Stra\ss e 38, 
D-01187 Dresden, Germany}
\ead{eckardt@pks.mpg.de}
\author{Egidijus Anisimovas}
\address{Department of Theoretical Physics, Vilnius University, Saul\.{e}tekio 9, LT-10222 Vilnius, Lithuania}
\address{Institute of Theoretical Physics and Astronomy, Vilnius University, Go\v{s}tauto 12, LT-01108 Vilnius, Lithuania}
\date{February 22, 2015}

\begin{abstract}
We derive a systematic high-frequency expansion for the effective Hamiltonian and the micromotion operator of periodically driven quantum systems. Our approach is based on the block diagonalization of the quasienergy operator in the extended Floquet Hilbert space by means of degenerate perturbation theory. The final results are equivalent to those obtained within a different approach [Phys.\ Rev.\ A {\bf 68}, 013820 (2003), Phys.\ Rev.\ X {\bf 4}, 031027 (2014)] and can also be related to the Floquet-Magnus expansion [J.\ Phys.\ A {\bf 34}, 3379 (2000)]. We discuss that the dependence on the driving phase, which plagues the latter, can lead to artifactual symmetry breaking. The high-frequency approach is illustrated using the example of a periodically driven Hubbard model. Moreover, we discuss the nature of the approximation and its limitations for systems of many interacting particles.  
\end{abstract}

\maketitle

\section{\label{sec:introduction} Introduction}
In the last years the concept of Floquet engineering has gained more and more interest. This form of quantum 
engineering is based on the fact that the time evolution of a periodically driven quantum system is, apart 
from a micromotion described by a time-periodic unitary operator, governed by a time-independent effective 
Hamiltonian \cite{Shirley65,Sambe73}. The aim is to engineer the properties of the effective Hamiltonian by 
designing a suitable time-periodic driving protocol.  
This concept has been employed very successfully in various experiments with ultracold atoms in driven optical 
lattices. This includes dynamic localization \cite{DunlapKenkre86,GrossmannEtAl91,Holthaus92,GrifoniHaenggi98,
LignierEtAl07,KierigEtAl08,EckardtEtAl09,CreffieldEtAl10}, ``photon''-assisted tunneling \cite{Zak93,
EckardtHolthaus07,SiasEtAl08,IvanovEtAl08,AlbertiEtAl09,AlbertiEtAl10,HallerEtAl10,MaEtAl11}, the control of
the bosonic superfluid-to-Mott-insulator transition \cite{EckardtEtAl05b,ZenesiniEtAl09}, resonant coupling of
Bloch bands \cite{GemelkeEtAl05,BakrEtAl11,ParkerEtAl13,HaEtAl15}, the dynamic creation of kinetic 
frustration \cite{EckardtEtAl10,StruckEtAl11}, as well as the realization of artificial magnetic fields and 
topological band structures \cite{OkaAoki09,Kolovsky11,BermudezEtAl11,AidelsburgerEtAl11,StruckEtAl11,
StruckEtAl12,HaukeEtAl12b,StruckEtAl13,AidelsburgerEtAl13,MiyakeEtAl13,AtalaEtAl14,AidelsburgerEtAl15,
JotzuEtAl14} (see also Ref.~\cite{RechtsmanEtAl13} for the creation of a topological band structure in an 
array of optical wave guides). In a quantum gas without a lattice, periodic driving has recently also been
employed to tune \cite{JimenezGarciaEtAl15} or induce \cite{LuoEtAl15} spin orbit coupling.

A prerequisite for Floquet engineering is a theoretical method to compute the effective Hamiltonian (as well 
as the micromotion operator), at least within a suitable approximation. In the high-frequency limit a
rotating-wave-type approximation can be employed for this purpose. This approximation coincides with the 
leading order of a systematic high-frequency expansion that provides also higher-order corrections
to the effective Hamiltonian and the micromotion operator 
\cite{RahavEtAl03,GoldmanDalibard14,ItinKatsnelson14,GoldmanEtAl15}. In this 
paper we show that this high-frequency expansion can be obtained  by employing degenerate 
perturbation theory in the extended Floquet Hilbert space. Our approach provides an intuitive picture of the 
nature of the approximation and the conditions under which it can be expected to provide a suitable description of a 
driven quantum system.
We point out that the time scale on which the approximation is valid can be increased by increasing the order of the 
approximation for the effective Hamiltonian, while keeping a lower-order approximation for the time-periodic micromotion 
operator. 
We also address the relation between the high-frequency expansion derived here and the Floquet-Magnus 
expansion \cite{CasasEtAl00} (see also Refs.~\cite{BlanesEtAl09,BukovEtAl14,VerdenyEtAl13}).
The origin of a spurious dependence of the quasienergy spectrum in Floquet-Magnus approximation on the driving phase is 
discussed (see also references \cite{RahavEtAl03,GoldmanDalibard14,GoldmanEtAl15}). Using the example of a circularly 
driven tight-binding lattice, this artifact is, moreover, shown to produce a non-physical breaking of the rotational 
symmetry in the approximate quasienergy band structure. Finally, we discuss the validity of the high-frequency approximation for systems of many interacting particles.

This paper is organized as follows. Section~\ref{sec:Floquet} gives a brief introduction to the theory of 
periodically driven quantum systems (Floquet theory) and serves to define our notation. In
Section~\ref{sec:block} we formulate the problem that is then attacked in Section~\ref{sec:HF} by means of the 
degenerate perturbation theory developed in \ref{sec:perturbation}. The relation to the Floquet-Magnus 
expansion is discussed in Section~\ref{sec:FM} and Section~\ref{sec:example} illustrates the approximation 
scheme using the example of a circularly driven hexagonal lattice \cite{OkaAoki09,RechtsmanEtAl13,JotzuEtAl14}.
Finally Section~\ref{sec:interactions} discusses effects of interactions within and beyond the high-frequency 
approximation, before we close with a brief summary in Section \ref{sec:conclusions}.

\section{\label{sec:Floquet}Quantum Floquet theory and notation}

\subsection{Floquet states}
A quantum system described by a time-periodic Hamiltonian 
\be
\Ho(t)=\Ho(t+T)
\ee
possesses generalized stationary states $|\psi_n(t)\ra$ called Floquet states \cite{Shirley65}. These 
states are solutions to the time-dependent Schr\"odinger equation
\be\label{eq:Schroedinger}
i\hbar\rd_t |\psi(t)\ra=\Ho(t)|\psi(t)\ra
\ee
of the form 
\be
|\psi_n(t)\ra=|u_n(t)\ra e^{-i\varepsilon_n t/\hbar},
\ee
with real \emph{quasienergy} $\varepsilon_n$ and time-periodic \emph{Floquet mode}
\be
|u_n(t)\ra=|u_n(t+T)\ra.
\ee
Here $\rd_t$ denotes the derivative with respect to the time $t$. The existence of Floquet states in
time-periodically driven systems follows from Floquet's theorem in a similar way as the existence of 
Bloch states in spatially periodic systems. For completeness, we give a simple proof for the existence 
of Floquet states in \ref{sec:existence}.

The Floquet states are eigenstates of the time-evolution operator over one driving period,
\be
\Uo(t_0+T,t_0)|\psi_n(t_0)\ra=e^{-i\varepsilon_n T/\hbar}|\psi_n(t_0)\ra. 
\ee
Here $\Uo(t_2,t_1)$
denotes the time evolution operator from time $t_1$ to time $t_2$. 
The eigenvalue $e^{-i\varepsilon_n T/\hbar}$ does not depend on the time
$t_0$ from which the evolution over one driving period starts. Therefore, one can obtain the quasienergy 
spectrum by computing and diagonalizing $\Uo(t_0+T,t_0)$ for an arbitrary $t_0$. The time-dependent
Floquet states $|\psi_n(t)\ra$ can subsequently be computed by applying the time-evolution operator, 
$|\psi_n(t)\ra=\Uo(t,t_0)|\psi_n(t_0)\ra$.

The Floquet states can be chosen to form a complete orthonormal basis at any fixed time $t$. As a 
consequence, the time evolution operator can be written like
\be
\Uo(t_2,t_1) 
=       \sum_n e^{-i\varepsilon_n(t_2-t_1)/\hbar}|u_n(t_2)\ra\la u_n(t_1)| .
\ee
Moreover, one can express the time evolution of a state $|\psi(t)\ra$ like 
\be\label{eq:TimeEvolution}
|\psi(t)\ra = 
    \sum_n  c_n  e^{-i\varepsilon_n (t-t_0)/\hbar} |u_n(t)\ra, 
\ee
with time-independent coefficients $c_n=\la u_n(t_0)|\psi(t_0)\ra$.
That is, if the system is prepared in a single Floquet state, $|c_n|=\delta_{n,n_0}$, its time evolution will be periodic
and (apart from the irrelevant overall phase factor $e^{-i\varepsilon_{n_0} t/\hbar}$) described by the Floquet mode
$|u_{n_0}(t)\ra$. If the system is prepared in a coherent superposition of several Floquet states, the time evolution 
will not be periodic anymore and be determined by two contributions. The first contribution stems from the periodic time 
dependence of the Floquet modes $|u_n(t)\ra$ and is called \emph{micromotion}. The second contribution, which leads to 
deviations from a periodic evolution, originates from the relative dephasing of the factors $e^{-i\varepsilon_n t/\hbar}$.
Thus, beyond the periodic micromotion, the time evolution of a Floquet system is governed by the quasienergies
$\varepsilon_n$ of the Floquet states in a similar way as the time evolution of an autonomous system 
(with time-independent Hamiltonian) is governed by the energies of the stationary states.

\subsection{Floquet Hamiltonian and micromotion operator}

In order to study the dynamics over time spans that are long compared to a single driving period, one 
can ignore the micromotion by studying the time evolution in a stroboscopic fashion in 
steps of the driving period $T$. Such a stroboscopic time evolution is described by the
time-independent \emph{Floquet Hamiltonian} $\Ho^F_{t_0}$. It is defined such that it generates the time evolution over one period, 
\be\label{eq:DefHeff}
\exp\bigg(-\frac{i}{\hbar}T\Ho^F_{t_0}\bigg)\equiv\Uo(t_0+T,t_0).
\ee
and can be expressed like
\be\label{eq:HeffModes}
\Ho^F_{t_0} =\sum_n \varepsilon_n |u_n(t_0)\ra\la u_n(t_0)| .
\ee
The parametric dependence on the initial time $t_0$ is periodic, $\Ho^F_{t_0+T}=\Ho^F_{t_0}$, and related to the 
micromotion. It indicates when during the driving period the dynamics sets in or is looked at and should not be 
confused with a time dependence of the Floquet Hamiltonian. 
From a Floquet Hamiltonian $\Ho^F_{t_0}$ obtained for the initial time $t_0$ one can construct a Floquet 
Hamiltonian for a different initial time $t_0'$ by applying a unitary transformation,
$\Ho^F_{t_0'}=\Uo^\dag(t_0,t_0')\Ho^F_{t_0}\Uo(t_0,t_0')$.

It is convenient to introduce a unitary operator that describes the periodic
time dependence of the Floquet modes, i.e.\ the micromotion. Such a two-point \emph{micromotion operator} can be
defined by  
\be\label{eq:Umicro}
\Uo_F(t_2,t_1) \equiv \sum_n |u_n(t_2)\ra\la u_n(t_1)|
\ee
so that, by construction, it evolves the Floquet modes in time,
\be
|u_n(t_2)\ra=\Uo_F(t_2,t_1)|u_n(t_1)\ra.
\ee
It is periodic in both arguments, $\Uo_F(t_2+T,t_1)=\Uo_F(t_2,t_1+T)=\Uo_F(t_2,t_1)$.

If the Floquet states and their quasienergies are known, e.g.\ from computing and diagonalizing the 
time evolution operator over one period, one can immediately write down the Floquet 
Hamiltonian and the micromotion operator by making use of Eqs.~(\ref{eq:HeffModes}) and
(\ref{eq:Umicro}). 
However, both the Floquet Hamiltonian $\Ho^F_{t_0}$ and the micromotion operator $\Uo_F(t,t')$ might also be 
computed directly, without computing the Floquet states and the quasienergies before. This will be the aim of the 
approximation scheme described in the main part of this paper. From the Floquet Hamiltonian and the micromotion operator 
one can then immediately write down the time evolution operator like
\be\label{eq:Evolution}
\Uo(t_2,t_1) 
=e^{-i(t_2-t_1) \Ho^F_{t_2 }/\hbar}\Uo_F(t_2,t_1)
=\Uo_F(t_2,t_1)e^{-i(t_2-t_1) \Ho^F_{t_1 }/\hbar}.
\ee
Moreover, the Floquet modes $|u_n(t_0)\ra$ and their quasienergies $\varepsilon_n$ can, in a subsequent 
step, be obtained from the diagonalization of $\Ho^F_{t_0}$, 
\be\label{eq:HeffEig}
\Ho^F_{t_0}|u_n(t_0)\ra=\varepsilon_n|u_n(t_0)\ra .
\ee
The periodic time-dependence of the Floquet modes can subsequently be computed by employing the 
micromotion operator, $|u_n(t)\ra=\Uo_F(t,t_0)|u_n(t_0)\ra$.

\subsection{Quasienergy eigenvalue problem and extended Floquet Hilbert space}
\label{sec:QOp}
The phase factors $e^{-i\varepsilon_n T/\hbar}$ and the Floquet states $|\psi_n(t)\ra$, solving the 
eigenvalue problem of the time-evolution operator over one period, are uniquely defined (apart from 
the freedom to multiply each Floquet state by a time independent phase factor). In turn, the 
quasienergies $\varepsilon_n$, and with them also the Floquet modes
$|u_n(t)\ra=e^{i\varepsilon_n t/\hbar}|\psi_n(t)\ra$ and the Floquet Hamiltonian (\ref{eq:HeffModes}),
are not defined uniquely. Namely, 
adding an integer multiple of $\hbar\omega$ to the quasienergy $\varepsilon_n$ does not alter the phase 
factor $e^{-i\varepsilon_n T/\hbar}$. Fixing each quasienergy $\varepsilon_n$ within this freedom fixes 
also the Floquet modes and the Floquet Hamiltonian. For example, one can choose all quasienergies to 
lie within the same interval of width $\hbar\omega$, often called a \emph{Brillouin zone}. This term reflects a loose 
analogy to the theory of spatially periodic Hamiltonians, where the quasimomentum can be chosen to lie within a single 
reciprocal lattice cell such as the first Brillouin zone.

Starting from the known solution given by $|u_n(t)\ra$ and $\varepsilon_n$, one can label all possible 
choices for the quasienergy by introducing the integer index $m$, 
\be\label{eq:epsper}
\varepsilon_{nm}=\varepsilon_n+m\hbar\omega.
\ee
The corresponding Floquet mode reads
\be\label{eq:uper}
 |u_{nm}(t)\ra = |u_n(t)\ra e^{im\omega t},
\ee
such that
\be\label{eq:decomposition}
|\psi_n(t)\ra = |u_{n}(t)\ra e^{-i\varepsilon_{n} t/\hbar} 
=|u_{nm}(t)\ra e^{-i\varepsilon_{nm} t/\hbar} .
\ee

When entering the right-hand side of Eq.~(\ref{eq:decomposition}) into the time-dependent Schr\"odinger 
equation (\ref{eq:Schroedinger}), we arrive at
\be\label{eq:PreQuasienergy}
[\Ho(t)-i\hbar\rd_t ]|u_{nm}(t)\ra = \varepsilon_{nm}|u_{nm}(t)\ra .
\ee
This equation constitutes an eigenvalue problem in an extended Hilbert space
$\mathcal{F}=\mathcal{H}\otimes\mathcal{L}_T$ \cite{Shirley65,Sambe73}. This space is given by the product space of the 
state space $\mathcal{H}$ of a quantum system and the space of square-integrable $T$-periodically time-dependent
functions $\mathcal{L}_T$. Time is treated as a coordinate under periodic boundary conditions. In the extended Floquet 
Hilbert space $\mathcal{F}$, the scalar product combines the scalar product of $\mathcal{H}$ with time averaging and is 
defined by
\be
\lla u|v\rra =\frac{1}{T}\int_0^T\!\rd t\, \la u(t)|v(t)\ra .
\ee 
We will use a double ket notation $|u\rra$ for elements of $\mathcal{F}$; the corresponding state at time $t$ in
$\mathcal{H}$ will be denoted by $|u(t)\ra$. Vice versa, a state $|v(t)\ra=|v(t+T)\ra$, including its full periodic time
dependence, is denoted by $|v\rra$ when considered as element of $\mathcal{F}$. In the following we will stick to this 
convention and conveniently switch between both representations. Likewise, an operator acting in $\mathcal{F}$ 
will be indicated by an overbar to distinguish it from operators acting in $\mathcal{H}$, which are 
marked by a hat. For example, $\bar{Q}$ denotes the $\mathcal{F}$-space operator that in
$\mathcal{H}$ is represented by
\be\label{eq:Q}
\hat{Q}(t) = \hat{H}(t)-i\hbar\rd_t.
\ee
The operator $\bar{Q}$ is called quasienergy operator. It is hermitian in $\mathcal{F}$ and, as can be 
inferred from Eq.~(\ref{eq:PreQuasienergy}), its eigenstates and eigenvalues are the Floquet modes and 
their quasienergies, 
\be\label{eq:qeigen}
\bar{Q} |u_{nm}\rra =\varepsilon_{nm}|u_{nm}\rra.
\ee
The complete set of solutions of the quasienergy eigenvalue problem (\ref{eq:qeigen}) contains a lot of 
redundant information. In the extended space $|u_{nm}\rra$ and $|u_{nm'}\rra$ constitute independent 
orthogonal solutions if $m'\ne m$. These solutions are, however, related to each other by
Eqs.\  (\ref{eq:epsper}) and (\ref{eq:uper}), and give rise to the same Floquet state $|\psi_n(t)\ra$.
All Floquet states $|\psi_n(t)\ra$ of the system can, thus, be constructed, e.g., from those Floquet 
modes whose quasienergies lie in a single Brillouin zone of the $\hbar\omega$-periodic quasienergy 
spectrum.

\begin{figure}[t]
\includegraphics[width=0.7\linewidth]{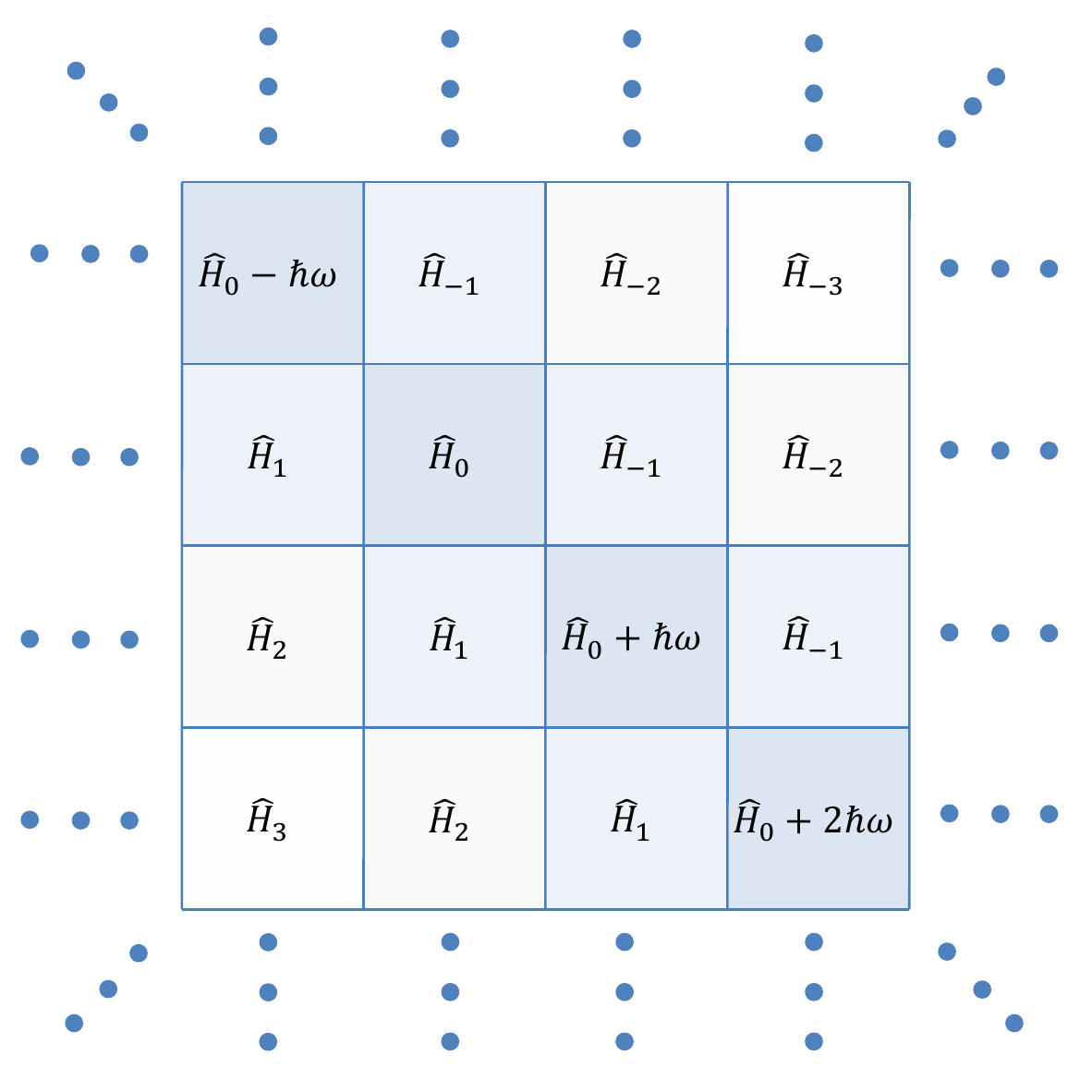}
\centering
\caption{\label{fig:quasienergy}Block structure of the quasienergy operator $\bar{Q}$ with respect to 
the ``photon'' index $m$. 
Each block corresponds to an operator $\Qo_{m'm}=\Ho_{m'-m}+\delta_{m'm}m\hbar\omega$ acting in the 
full state space $\mathcal{H}$. The diagonal blocks $\Ho_0+m\hbar\omega$ can be interpreted to act in the 
subspace of relative ``photon'' number $m$ and the off-diagonal blocks $\Ho_{m'-m}$, which obey
$\Ho_{m'-m}=\Ho^\dag_{m-m'}$, describe $(m'-m)$-``photon'' processes.}
\end{figure}

The quasienergy eigenvalue problem (\ref{eq:qeigen}) provides a second approach for computing the 
Floquet states or the Floquet Hamiltonian, alternative to the computation and diagonalization of the 
time evolution operator over one driving period. It provides the Floquet modes not only at a time $t_0$, 
but including their full periodic time dependence. Despite the drastically increased Hilbert space, 
treating the quasienergy eigenvalue problem (\ref{eq:qeigen}) has also advantages. In order to 
diagonalize the hermitian quasienergy operator $\bar{Q}$, one can employ methods, concepts, and 
intuition from the physics of systems with time-independent Hamiltonians. When describing parameter
variations, such as a smooth switching on of the driving amplitude, one can even derive a
Schr\"odinger-type evolution equation acting in Floquet space and apply the adiabatic principle
\cite{BreuerHolthaus89b}\footnote{A discussion of smooth parameter
variations in a driven many-body lattice system can be found in reference \cite{EckardtHolthaus08b}.}.

A complete set of orthonormal basis states $|\alpha m\rra$ of $\mathcal{F}$ can be 
constructed by combining a complete set of orthonormal basis states $|\alpha\ra$ of $\mathcal{H}$
with the complete set of time-periodic functions $e^{im\omega t}$ labeled by the integer $m$,
\be\label{eq:simplebasis}
|\alpha m(t)\ra = |\alpha\ra e^{im\omega t}.
\ee
From this restricted class of basis states all possible sets of basis states can now be constructed by 
applying unitary operators $\bar{U}$, $|\alpha m\rra_U=\bar{U}|\alpha m\rra$.
With respect to the basis $|\alpha m\rra$ the quasienergy operator possesses the matrix elements
\bes\label{eq:me}
\lla \alpha'm'|\bar{Q}|\alpha m\rra
&=&\frac{1}{T}\int_0^T \!\rd t\, e^{-i m'\omega t}\la\alpha'|\Ho(t)-i\hbar\rd_t|\alpha\ra e^{i m\omega t}
\nonumber\\
&=& \la\alpha'|\Ho_{m'-m}|\alpha\ra+\delta_{m'm}\delta_{\alpha'\alpha}m\hbar\omega,
\ees
where
\be
\Ho_{m} = \frac{1}{T}\int_0^T \! \rd t\, e^{-i m\omega t}\Ho(t) =\Ho_{- m}^\dag
\ee
is the Fourier transform of the Hamiltonian $\Ho(t)$, such that
$\Ho(t)=\sum_{ m=-\infty}^\infty e^{i m\omega t}\Ho_{m}$.
With respect to the Fourier indices $m$ the quasienergy operator possesses the transparent block
structure depicted in Fig.~\ref{fig:quasienergy}. Each block represents an operator
$\Qo_{m'm}=\Ho_{m'-m}+\delta_{m'm}m\hbar\omega$ acting in $\mathcal{H}$.

The structure of the quasienergy operator $\bar{Q}$ resembles that of the Hamiltonian describing a 
quantum system with Hilbert space $\mathcal{H}$ coupled to a photon-like mode in the classical limit of
large photon numbers, where the spectrum becomes periodic in energy. In this picture $m$ plays the role 
of a relative photon number. The quasienergy eigenvalue problem (\ref{eq:qeigen}) is, thus, closely 
related to the dressed-atom picture \cite{AtomPhotonInteractions,AvanEtAl76} for a quantum system driven by coherent 
radiation \cite{EckardtHolthaus08a}. Based on this analogy, one often uses the jargon to call
$m$ the ``photon'' number. 
Moreover, the matrix elements of $\Ho_{m}$ are said to describe $ m$-``photon'' processes. This 
terminology suggests a very intuitive picture for the physics of time-periodically driven quantum systems and
is also employed when the system is actually not driven by a photon mode.

In order to diagonalize or block diagonalize the quasienergy operator, it is natural and sufficient to consider unitary 
operators $\bar{U}$ that are translationally invariant with respect to the photon index $m$, 
$\lla \alpha'm'|\bar{U}|\alpha m\rra= \la\alpha'|\Uo_{m'-m}|\alpha\ra$. They correspond to
time-periodic unitary operators $\Uo(t)=\sum_{m=-\infty}^\infty e^{i m\omega t}\Uo_{ m} $ acting in $\mathcal{H}$
(see also \ref{sec:operators}). 
From Eq.~(\ref{eq:Q}) we can infer that a unitary transformation with such an operator $\bar{U}$, 
\bes\label{eq:Funitary}
\bar{Q}&\to&\bar{Q}'=\bar{U}^\dag\bar{Q}\bar{U}
\nonumber\\
|u\rra&\to&|u'\rra=\bar{U}^\dag|u\rra,
\ees
is equivalent to a gauge transformation
\bes\label{eq:Hgauge}
\Ho(t)&\to&\Ho'(t)=\Uo^\dag(t)\Ho(t)\Uo(t)-i\hbar\Uo^\dag(t)\rd_t\Uo(t)
\nonumber\\
|\psi(t)\ra&\to&|\psi'(t)\ra=\Uo^\dag(t)|\psi(t)\ra.
\ees
with a time-periodic unitary operator $\Uo(t)$. Accordingly, the matrix elements of the transformed quasienergy operator
\be
\lla \alpha'm'|\bar{Q}'|\alpha m\rra 
= \la\alpha'|\Ho_{m'-m}'|\alpha\ra+\delta_{m'm}\delta_{\alpha'\alpha}m \hbar\omega,
\ee
are determined by the Fourier components $\Ho_m'=\frac{1}{T}\int_0^T\!\rd t\,e^{-im\omega t}\Ho'(t)$ of the
gauge-transformed Hamiltonian. 

The unitary operator $\bar{U}_D$ that diagonalizes the quasienergy operator with respect to a 
certain basis $|\alpha m\rra$, 
\be
\lla \alpha'm'|\bar{U}_D^\dag\bar{Q}\bar{U}_D|\alpha m\rra 
= \delta_{m'm}\delta_{\alpha'\alpha} (\la\alpha|\Ho_D|\alpha\ra + m \hbar\omega)
\ee
is constructed such that it leads to a time-independent gauge-transformed Hamiltonian
\be
\Ho_D =\Uo_D^\dag(t)\Ho(t)\Uo_D(t)-i\hbar\Uo_D^\dag(t)\rd_t\Uo_D(t)
\ee
that is diagonal with respect to the basis states $|\alpha\ra$, 
\be
\la\alpha'|\Ho_D|\alpha\ra =\delta_{\alpha'\alpha}\varepsilon_\alpha.
\ee  
The Floquet Hamiltonian $\Ho^F_{t_0}$ is related to $H_D$ via the unitary transformation
\be
\Ho^F_{t_0} = \Uo_D(t_0) \Ho_D \Uo_D^\dag(t_0).
\ee
The Floquet mode $|u_{\alpha m}\rra=\bar{U}_D|\alpha m\rra$ with quasienergy
$\varepsilon_{\alpha m}=\varepsilon_\alpha+m\hbar\omega$ reads
$|u_{\alpha m}(t)\ra = \Uo_D(t)|\alpha\ra e^{im\omega t}$, so that the micromotion operator can be expressed like 
\be
\Uo_F(t,t')=\Uo_D(t)\Uo_D^\dag(t').
\ee

\section{\label{sec:block}Block diagonalization of the quasienergy operator and effective Hamiltonian}

\begin{figure}[t]
\includegraphics[width=0.7\linewidth]{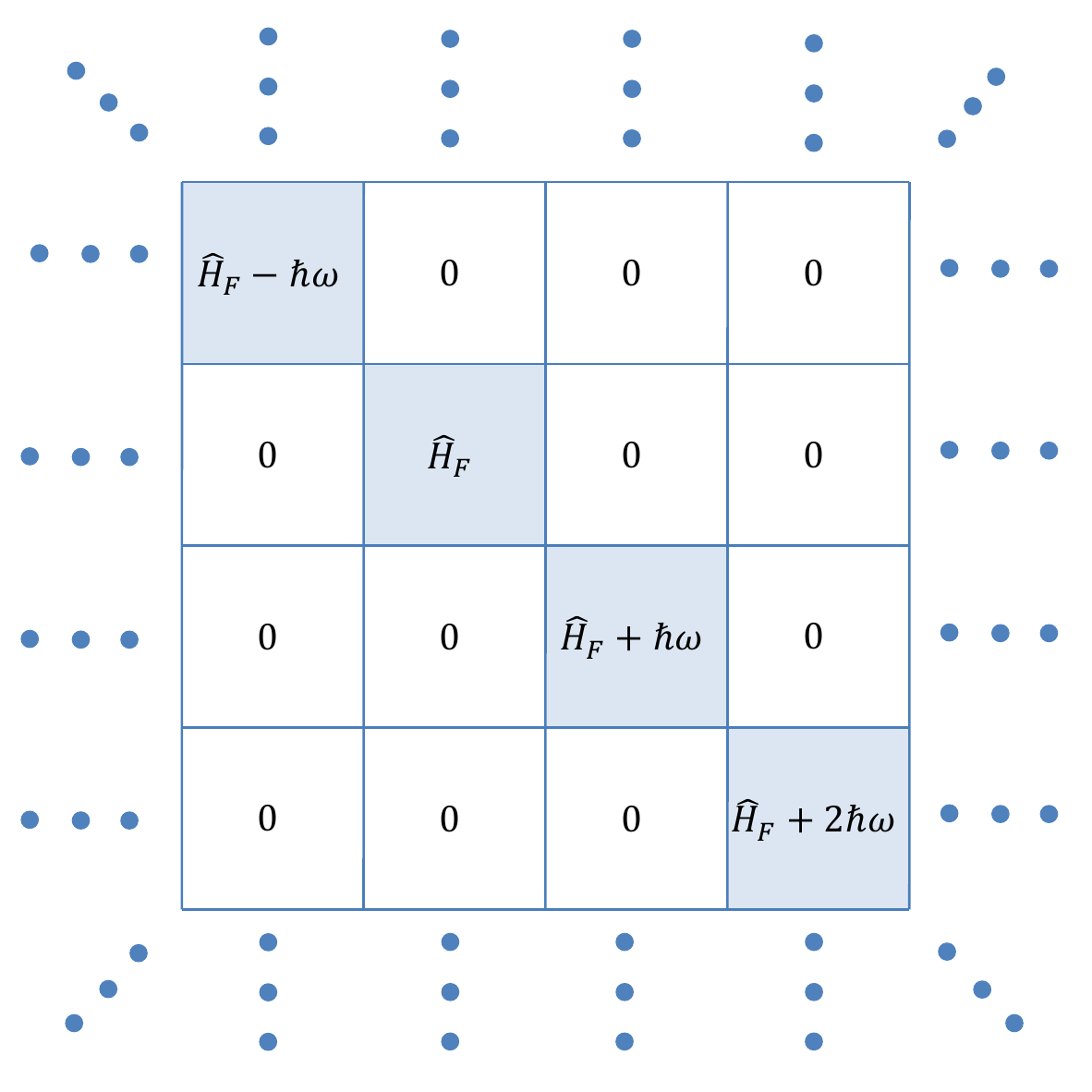}
\centering
\caption{\label{fig:H_F}By block diagonalization of the quasienergy operator with respect to the photon index $m$, one obtains the effective Hamiltonian $\Ho_F$.}
\end{figure}

The quasienergy eigenvalue problem (\ref{eq:qeigen}) is a convenient starting point for computing the
Floquet Hamiltonian and the micromotion operator directly, without the need to compute the 
Floquet modes and their quasienergies. For this purpose one does not need to fully diagonalize the quasienergy operator. 
Instead one has to find a unitary operator $\bar{U}_F$ that \emph{block diagonalizes} the quasienergy operator with 
respect to the ``photon'' index $m$,
\be\label{eq:HFloquet}
\lla\alpha'm'|\bar{U}^\dag_F\bar{Q}\bar{U}_F|\alpha m\rra 
= \delta_{m'm} (\la\alpha'|\Ho_F|\alpha\ra+\delta_{\alpha'\alpha}m\hbar\omega),
\ee
as illustrated in Fig.~\ref{fig:H_F}. 
Here we have introduced the gauge-transformed Hamiltonian
\be\label{eq:BlockGauge}
\Ho_{F}=\Uo^\dag_F(t)\Ho(t)\Uo_F(t)-i\hbar\Uo^\dag_F(t)\rd_t\Uo_F(t),
\ee
which by construction is time independent. In fact, choosing an operator $\bar{U}_F$ that block diagonalizes the 
quasienergy operator is equivalent to choosing $\Uo_F(t)$ such that the gauge transformation (\ref{eq:BlockGauge}) leads 
to a time-independent Hamiltonian $\Ho_F$. This time-independent Hamiltonian $\Ho_F$ is called
\emph{effective Hamiltonian}. Note that the unitary operator $\bar{U}_F$ is not determined uniquely. For example, 
multiplying $\Uo_F(t)$ with any time-independent unitary operator from the right leads to a mixing of states within the 
diagonal blocks of $\bar{U}^\dag_F\bar{Q}\bar{U}_F$, but does not destroy the block diagonal form. Unlike the operator
$\bar{U}_D$ that diagonalizes the quasienergy operator, $\bar{U}_F$ does not depend on the basis states $|\alpha\ra$.

Now, each of the diagonal blocks of $\bar{U}^\dag_F\bar{Q}\bar{U}_F$ represents a possible choice for the Floquet 
Hamiltonian. This can be seen by writing the quasienergy operator like
$\bar{Q}=\sum_{m=-\infty}^\infty\sum_n|u_{nm}\rra (\varepsilon_\alpha +m\hbar\omega)\lla u_{nm}|$ and comparing it to the 
representation (\ref{eq:HeffModes}) of the Floquet Hamiltonian.
From the $m=m'=0$ block one obtains 
\bes\label{eq:BloDi}
\Ho^F_{t_0} 
&=&\sum_{\alpha'\alpha} 
    |\alpha'0(t_0)\ra_{F}\,{}_{F}\lla\alpha' 0|\bar{Q}|\alpha 0\rra_{F}\,{}_{F}\la\alpha 0(t_0)|
\nonumber\\
&=&\sum_{\alpha'\alpha} \Uo_F(t_0)|\alpha'\ra\,\lla\alpha' 0|\bar{U}_F^\dag\bar{Q}\bar{U}_F
                |\alpha 0\rra\,\la\alpha|\Uo_F^\dag(t_0)
\nonumber\\
&=& \Uo_F(t_0) \Ho_F\Uo_F^\dag(t_0),
\ees
where we have defined the rotated basis states 
\be
|\alpha m\rra_F\equiv\bar{U}_F|\alpha m\rra
\ee
and used Eq.~(\ref{eq:HFloquet}). We can see that the Floquet Hamiltonian $\Ho^F_{t_0}$ is equivalent to the effective 
Hamiltonian $\Ho_F$ in the sense that both are related to each other by a unitary transformation. 
Moreover, we can use the unitary operator $\bar{U}_F$ to construct the micromotion operator: 
\be\label{eq:UbUmicro}
\Uo_F(t,t')=\Uo_F(t)\Uo_F^\dag(t').
\ee
From $\Ho^F_{t}$ and $\Uo_F(t,t')$ one can then directly obtain the time evolution operator using
Eq.~(\ref{eq:Evolution}). 

However, the time evolution operator $\Uo(t_2,t_1)$ can also be expressed directly 
in terms of $\Ho_F$ and $\Uo_F(t)$ without introducing $\Ho^F_{t_0}$ and $\Uo_F(t',t)$. Namely \cite{RahavEtAl03}, 
\be\label{eq:UHFUF}
\Uo(t_2,t_1)=\Uo_F(t_2)e^{-\frac{i}{\hbar}(t_2-t_1)\Ho_F}\Uo_F^\dag(t_1).
\ee
Compared to the representation (\ref{eq:Evolution}) of the time-evolution operator in terms of the Floquet Hamiltonian
$\Ho^F_{t_0}$ and the micromotion operator $\Uo_F(t_2,t_1)$, this expression has the disadvantage that it is a 
product of three operators and not just of two. However, using the representation (\ref{eq:UHFUF}) has also advantages.  
The micromotion has been expressed by the one-point micromotion operator $\Uo_F(t)$, instead of by the two-point operator 
$\Uo_F(t,t')$, and the phase evolution is described by an effective Hamiltonian $\Ho_F$ without the parametric dependence 
on the switching time $t_0$ of $\Ho^F_{t_0}$. 
The micromotion operator $\Uo_F(t)$ can also be expressed like
\be
\Uo_F(t)=\exp\big(\hat{G}(t)\big)
\ee
in terms of an anti-hermitian operator $\hat{G}=-\hat{G}^\dag$. The hermitian operator 
$\hat{K}(t)=i\hat{G}(t)$ has recently been given the intuitive name \emph{kick operator}
\cite{GoldmanDalibard14}. 

The diagonalization of the effective Hamiltonian $\Ho_F$,
\be
\Ho_F|\tilde{u}_n\ra=\varepsilon_{n}|\tilde{u}_n\ra,
\ee
provides the Floquet modes and their quasienergies:
\bes
|u_{nm}(t)\ra &=& \Uo_F(t)|\tilde{u}_n\ra e^{im\omega t},
\\
 \varepsilon_{nm}&=&\varepsilon_n+m\hbar\omega.
\ees
Thus, the Floquet modes $|u_n(t)\ra\equiv|u_{n0}(t)\ra$, which describe the micromotion, are 
superpositions 
\be
|u_{n}(t)\ra =\sum_\alpha \gamma_{\alpha n} |\alpha(t)\ra_F.
\ee
of the time-dependent basis states
\be
|\alpha(t)\ra_F=\Uo_F(t)|\alpha 0\ra =\Uo_F(t)|\alpha\ra,
\ee
with time-independent coefficients 
\be
\gamma_{\alpha n}=\la\alpha|\tilde{u}_n\ra.
\ee

The strategy of computing the effective Hamiltonian directly, without computing the Floquet states before,
separates the Floquet problem into two distinct subproblems related to the short-time and the long-time dynamics, 
respectively. 
The first problem, computing the effective Hamiltonian (as well as the micromotion operator), concerns the
short-time dynamics within one driving period only. 
The second problem consists in the integration of the time evolution generated by the effective 
Hamiltonian for a given initial state or even in the complete diagonalization of the effective 
Hamiltonian. This separation allows to address the long-time dynamics over several driving periods in a very 
efficient way, without the need to follow the details of the dynamics within every driving period. 

The advantage of splitting the Floquet problem into two parts becomes apparent especially when one of 
the two problems is more difficult than the other. 
A simple example for a case where computing the effective Hamiltonian is more difficult than diagonalizing
it, is a periodically driven two-level system corresponding to a spin-1/2 degree of freedom. While the block 
diagonalization of the quasienergy operator can generally not be accomplished analytically, the effective
Hamiltonian describes (like every time-independent $2\times2$ Hamiltonian) a spin 1/2 in a constant magnetic
field leading to a simple precession dynamics on the Bloch sphere. Thus, once the effective Hamiltonian and
the micromotion operator are computed, the time evolution is known.  
An example for the opposite case, where the effective Hamiltonian can be computed at least approximately 
while its diagonalization is much harder, is a time-periodically driven Hubbard-type model \cite{EckardtEtAl05b}.
It describes interacting particles on a tigh-binding lattice. This driven model allows for a quantitative description of 
experiments with ultracold atoms in optical lattices. In the limit of high-frequency forcing a suitable
analytical approximation to the effective Hamiltonian can be well justified on the time scale of a typical
optical lattice experiment. However, the effective Hamiltonian will constitute a many-body problem 
that is difficult to solve.  

The possibility to compute the effective Hamiltonian for a many-body lattice system, at least within a 
suitable approximation, is also the basis for a novel and powerful type of quantum engineering. Here
the properties of the effective Hamiltonian $\Ho_F$ are tailored by engineering the periodic 
time dependence of the Hamiltonian $\Ho(t)$. This \emph{Floquet engineering} has recently been 
successfully applied to ultracold atomic quantum gases (see references in the introduction). 
The fact that the effective Hamiltonian can possess properties that are hard to achieve otherwise, like 
the coupling of the kinetics of charge-neutral atoms to a vector potential describing an (artificial) 
magnetic field \cite{AidelsburgerEtAl11,StruckEtAl12,AidelsburgerEtAl13,MiyakeEtAl13,
AtalaEtAl14,AidelsburgerEtAl15,Kolovsky11,DalibardEtAl11,GoldmanEtAl14,HaukeEtAl12b,OkaAoki09,
KitagawaEtAl10,KitagawaEtAl11,LindnerEtAl11,JotzuEtAl14,RechtsmanEtAl13,CayssolEtAl13}, 
makes Floquet engineering also interesting for quantum simulation. Here, a quantum mechanical many-body model is realized 
accurately in the laboratory in order to investigate its properties by doing experiments. An essential prerequiste for 
Floquet engineering is an accurate approximation to the effective Hamiltonian. In the next section we will systematically 
derive a high-frequency approximation to both the effective Hamiltonian and the micromotion operator by block 
diagonalizing the quasienergy operator by means of degenerate perturbation theory.

\section{\label{sec:HF}High-frequency expansion from degenerate perturbation theory}

Degenerate perturbation theory is a standard approximation scheme for the systematic block 
diagonalization of a hermitian operator into two subspaces---a subspace of special interest on the one 
hand and the rest of state space on the other---that are divided by a large spectral gap. Here we adapt 
the method such that it allows for a systematic block diagonalization of the quasienergy operator with 
respect to the ``photon'' index $m$ (\ref{sec:perturbation}).
Moreover, we will identify the system-independent ``photonic'' part $-i\hbar\rd_t$ of the quasienergy operator
(\ref{eq:Q}), with $\lla \alpha'm'|-i\hbar\rd_t|\alpha m\rra =\delta_{m'm}\delta_{\alpha'\alpha}m \hbar\omega$, as the 
unperturbed problem. As a consequence the system-specific Hamiltonian $\Ho(t)$ constitutes the perturbation. 
This will allow us to systematically derive simple and universal expansions for both the effective Hamiltonian $\Ho_F$ 
and the micromotion operator $\Uo_F(t)$ in the high-frequency limit, where $\hbar\omega$ constitutes a large spectral gap 
between the unperturbed subspaces (see Fig.~\ref{fig:quasienergy}). We would like to point out that the application of 
degenerate perturbation theory in the extended Floquet Hilbert space is a well established method. For example, it has 
recently been employed to estimate the matrix element for the resonant creation of collective excitations in a driven
Bose-Hubbard model \cite{EckardtHolthaus08b} and to treat a dissipative driven two-level system \cite{HausingerGrifoni10}.

The basic strategy of our perturbative approach can be summarized as follows. The quasienergy operator 
is divided into an unperturbed part $\bar{Q}_0$ and a perturbation~$\bar{V}$,
\be
\bar{Q}=\bar{Q}_0+\bar{V}.
\ee
The unperturbed operator can be diagonalized and separates the extended Floquet Hilbert space
$\mathcal{F}$ into uncoupled subspaces $\mathcal{F}^{(0)}_m$ of sharp ``photon'' numbers $m$ with 
projectors $\bar{P}_m=\sum_\alpha|\alpha m\rra\lla\alpha m|$. These subspaces shall be separated by unperturbed 
spectral gaps of the order of $\hbar\omega$, which are assumed to be large compared to the strength $p$
of the perturbation coupling states of different subspaces. When smoothly switching on the perturbation, such that 
the spectral gaps do not close, the unperturbed subspaces $\mathcal{F}^{(0)}_m$ will be transformed 
adiabatically to the perturbed subspaces $\mathcal{F}_m$ corresponding to a diagonal block of the 
perturbed problem. Since the perturbation is weak compared to the gap, $\mathcal{F}_m$ will differ from
$\mathcal{F}^{(0)}_m$ by small admixtures of states $\not\in \mathcal{F}^{(0)}_m$ only. This admixture 
will be calculated perturbatively by expanding a unitary operator $\bar{U}_F$ that relates the basis 
states $|\alpha m\rra$ spanning the unperturbed subpsaces $\mathcal{F}_m^{(0)}$ to the basis states
$|\alpha m\rra_F=\bar{U}_F|\alpha m\rra$ spanning the perturbed subspaces $\mathcal{F}_m$. 
In contrast, if the spectral gap separating different subspaces would close, arbitrary weak coupling can hybridize 
degenerate states of different subspaces, contrary to the assumption of a weak perturbative admixture. 
The general formalism is developed in \ref{sec:perturbation} and will be applied to a specific choice of the unperturbed 
problem in the following.

\begin{figure}[t]
\includegraphics[width=1\linewidth]{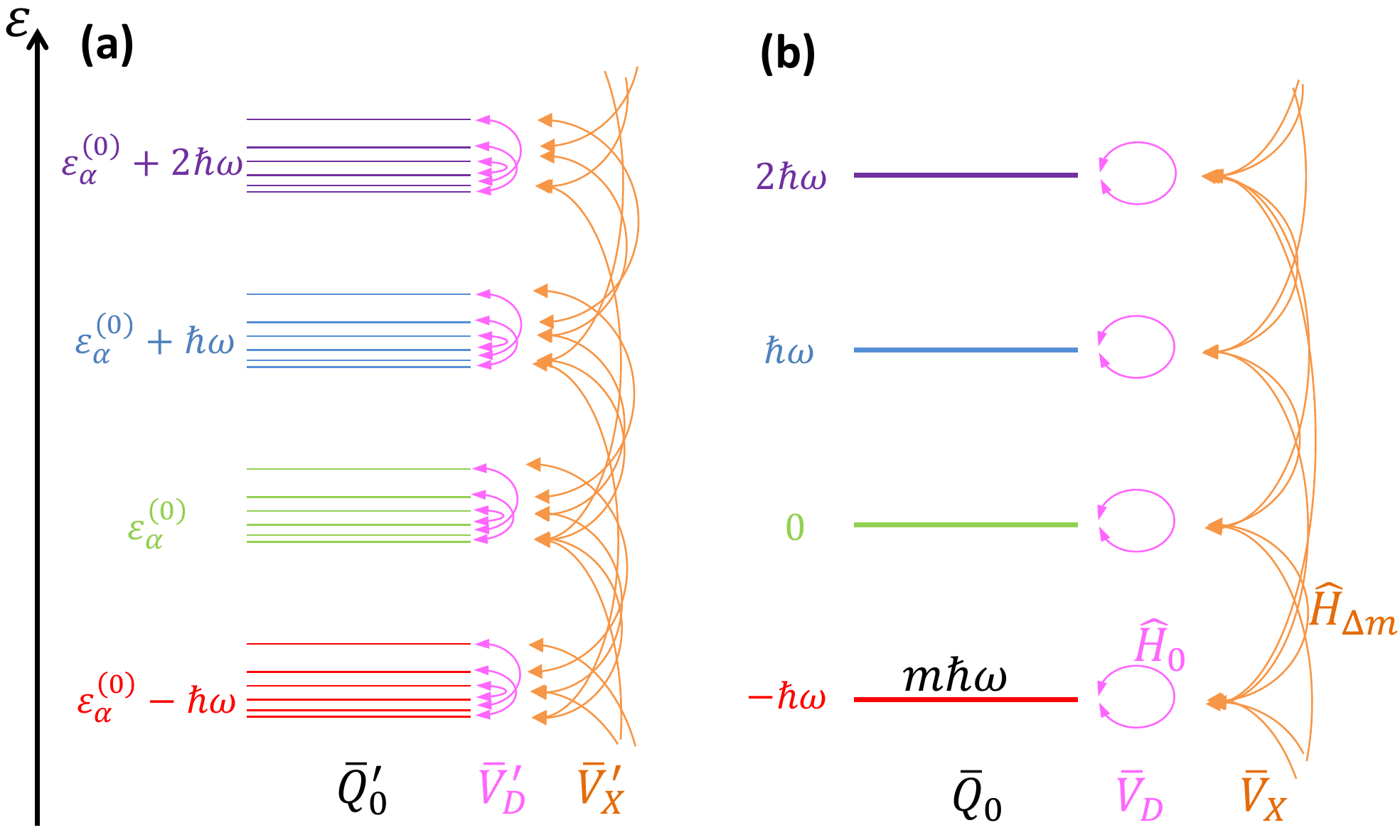}
\centering
\caption{\label{fig:QDeg}Structure of the quasienergy operator 
$\bar{Q}=\bar{Q}'_0+\bar{V}'_D+\bar{V}'_X=\bar{Q}_0+\bar{V}_D+\bar{V}_X$, the unperturbed problem is 
given by $\bar{Q_0}^{(')}$ and the perturbation $\bar{V}^{(')}$ can be separated into a block diagonal 
part $\bar{V}_D^{(')}$ that conserves the ``photon'' number $m$ and a part $\bar{V}_X^{(')}$ comprising
$m\ne0$-photon processes. (a) Generic choice of the unperturbed operator $\bar{Q}_0'$. (b) Simple
system-independent choice of the unperturbed problem $\Qo_0(t)= - i\hbar\rd_t$ to be used here; all 
unperturbed states of identical ``photon'' number $m$ are degenerate. }
\end{figure}

For the procedure described above a general and legitimate choice of the unperturbed problem would 
consist in the diagonal terms of the quasienergy operator with respect to a conveniently chosen set of 
basis states $|\alpha,m\rra$, 
\bes\label{eq:Q0NotUsed}
\bar{Q}'_0
    &=&\sum_m\sum_{\alpha} |\alpha m\rra \lla\alpha m|\bar{Q}|\alpha m\rra \lla\alpha m|
\nonumber\\
    &=& \sum_m\sum_{\alpha} |\alpha m\rra (\varepsilon_\alpha^{(0)}+m\hbar\omega) \lla\alpha m|,
\ees
with $\varepsilon_\alpha^{(0)}=\la\alpha|\Ho_0|\alpha\ra$. The operator $\bar{Q}'_0$ is diagonal with respect 
to the basis states $|\alpha m\rra$ by construction and the corresponding perturbation
$\bar{V}'=\bar{Q}-\bar{Q}'_0$ consists of a block-diagonal part $\bar{V}'_D$ that couples states
$|\alpha m\rra$ and $|\alpha'm\rra$ of the same ``photon'' number $m$ and a block-off-diagonal part
$\bar{V}'_X$ that couples states $|\alpha m\rra$ and $|\alpha'm'\rra$ of different ``photon'' numbers $m'$ 
and $m$. The problem to be solved by perturbation theory is visualized in Figure~\ref{fig:QDeg}(a). The
unperturbed problem and the perturbation expansion depend on the choice of the basis states $|\alpha\ra$. 

However, for the sake of simplicity 
we will not use Eq.~(\ref{eq:Q0NotUsed}). Instead we will simplify the unperturbed problem further, 
reducing it to the ``photonic'' part of the quasienergy operator, 
\be\label{eq:unperturbed}
\Qo_0(t)=-i\hbar\rd_t,
\ee
or
\be
\bar{Q}_0= \sum_m\sum_{\alpha} |\alpha m\rra m\hbar\omega \lla\alpha m|,
\ee
which does not depend on the system's Hamiltonian. For this choice the unperturbed quasienergies are 
degenerate within each subspace and read $\varepsilon_{\alpha m}^{(0)}=m\hbar\omega$. So $\bar{Q}_0$ is diagonal not 
only with respect to a specific set of basis states, but with respect to \emph{any} set of basis states of 
the type $|\alpha m\rra$. The perturbation is given by the Hamiltonian,
\be
\Vo(t)=\Ho(t)
\ee
or 
\be
\bar{V}=\bar{H}
    =\sum_{m'm}\sum_{\alpha'\alpha} |\alpha' m'\rra \la\alpha'|\Ho_{m'-m}|\alpha\ra \lla\alpha m|.
\ee
It can be decomposed like
\be
\bar{V}=\bar{V}_D+\bar{V}_X.
\ee
Here the block-diagonal part $\bar{V}_D$ comprises the $m'=m$ terms describing zero-``photon'' processes 
determined by the time-averaged Hamiltonian,
\bes
\Vo_D &=& \Ho_0,
\\
\bar{V}_D&=&\sum_{m}\sum_{\alpha'\alpha} |\alpha' m\rra \la\alpha'|\Ho_{0}|\alpha\ra \lla\alpha m|.
\ees
The block-off-diagonal part $\bar{V}_X$ describes $\Delta m$-``photon'' processes determined by the Fourier 
components $\Ho_{\Delta m}$ of the Hamiltonian,
\bes
\Vo_X(t) &=& \sum_{\Delta m\ne 0}e^{i \Delta m\omega t}\Ho_{\Delta m}
\\
\bar{V}_X&=&\sum_{m}\sum_{\Delta m\ne 0}\sum_{\alpha'\alpha} 
            |\alpha' m+\Delta m\rra \la\alpha'|\Ho_{\Delta m}|\alpha\ra \lla\alpha m|.
\ees
The problem is visualized in Figure~\ref{fig:QDeg}(b). Its simple structure will allow us to write down 
universal analytical expressions for the leading terms of a perturbative high-frequency expansion of the 
effective Hamiltonian and the micromotion operator in powers of $p/\hbar\omega$, with $p$ symbolizing 
the perturbation strength.

Before moving on, we note in passing that it can be useful to shift the ``photon'' number of an unperturbed state
$|\alpha m\rra$ by some integer $\Delta m$, before applying the high-frequency approximation. 
Such a procedure can be useful, if two states $|\alpha_1\ra$ and $|\alpha_2\ra$ have time-averaged energies
$\varepsilon^{(0)}_\alpha=\la\alpha|\Ho_0|\alpha\ra$ that are separated by $\Delta m$ ``photon'' energies of $\hbar\omega$,
so that $\varepsilon^{(0)}_{\alpha_2}-\varepsilon^{(0)}_{\alpha_1} =\Delta m\hbar\omega + \delta$ with
$\delta \ll \hbar\omega$. 
In this case the two unperturbed basis states $|\alpha_1 m\rra$ and $|\alpha_2 m\rra$, which are degenerate with respect 
to the unperturbed quasienergy operator, have average quasienergies 
$\lla\alpha m|\bar{Q}|\alpha m\rra=\varepsilon^{(0)}_\alpha+m\hbar\omega$ that are also separated by the large distance
$\Delta m\hbar\omega + \delta$. Obviously, this violates the requirement that the perturbation should be weak. In turn 
the states $|\alpha_1 m\rra$ and $|\alpha_2 (m-\Delta m)\rra$ have average quasienergies that are nearly degenerate. Thus 
it is useful to redefine the ``photon'' number of states with quantum number $\alpha_2$, so that 
$|\alpha_2 m\rra'=|\alpha_2 (m-\Delta m)\rra$. This redefinition is equivalent to a gauge transformation (\ref{eq:Hgauge}) 
in $\mathcal{H}$, where the unitary operator $e^{-i\Delta m\omega t}|\alpha_2\ra\la\alpha_2|$ is employed to shift the 
time-averaged energy of $|\alpha_2\ra$ by $-\Delta m\hbar\omega$. After this transformation the high-frequency 
approximation can be applied and used to describe the resonant coupling between both states $|\alpha_1\ra$ and
$|\alpha_2\ra$. Such a procedure can be employed, e.g., to describe resonant ``photon''-assisted (or AC-induced) 
tunneling against a strong potential gradient \cite{EckardtHolthaus07,GoldmanEtAl15}.

\subsection{Micromotion}
We wish to compute the unitary operator $\bar{U}_F$ that relates the unperturbed basis states $|\alpha m\rra$ 
to the perturbed basis states $|\alpha m\rra_F$ that block diagonalize the quasienergy operator in a 
perturbative fashion. In the canonical van Vleck degenerate perturbation theory, it is written like
\be\label{eq:AnsatzUB}
\bar{U}_F=\exp(\bar{G}),
\ee
with anti-hermitian operator
\be
\bar{G}=-\bar{G}^\dag.
\ee
In order to minimize the mixing of unperturbed states belonging to the same unperturbed subspace, it is, 
moreover, required that $\bar{G}$ is block off diagonal. One can now systematically expand $\bar{G}$ like
\be\label{eq:Gexp}
\bar{G}=\sum_{\nu=1}^\infty \bar{G}^{(\nu)}
\ee
in powers of the perturbation. The general formalism for the perturbative expansion of $\bar{G}$ in a 
situation where the state space is partitioned into more than just two subspaces is described in
\ref{sec:perturbation}. Differences with respect to the standard procedure, where the state space is just
bipartitioned, arise as a consequence of the fact that for multipartitioning it is generally not true 
anymore that the product of two block-off-diagonal operators is block diagonal. 

The general form of the leading terms of the expansion (\ref{eq:Gexp}) is given by Eqs.~(\ref{eq:G1}) 
and (\ref{eq:G2}) of \ref{sec:perturbation}. Let us evaluate them for the particular choice of the 
unperturbed problem (\ref{eq:unperturbed}). Apart from 
\be
\lla \alpha' m|\bar{G}^{(\nu)}|\alpha m\rra = 0 ,
\ee
for all diagonal matrix elements, following directly from $\bar{G}$ being block-off-diagonal, for $m'\ne m$ we obtain
\be\label{eq:G1p}
\lla \alpha' m'|\bar{G}^{(1)} |\alpha m\rra 
        =-\frac{\la \alpha'|\Ho_{m'-m}|\alpha\ra }{(m'-m) \hbar\omega}              
\ee
and
\bes\label{eq:G2p}
\lla \alpha' m'|\bar{G}^{(2)}|\alpha m\rra &=& 
\frac{\la\alpha'|\big[\Ho_0,\Ho_{m'-m}\big]|\alpha\ra}{[(m'-m)\hbar\omega]^2}
\nonumber\\&&+\,\frac{1}{2}
    \sum_{m''\ne m,m'}
    \frac{\la\alpha'|\Ho_{m'-m''}\Ho_{m''-m}|\alpha\ra}{(m'-m)\hbar\omega}
\nonumber\\&&\times         
\bigg[\frac{1}{(m''-m')\hbar\omega}+\frac{1}{(m''-m)\hbar\omega}\bigg].
\ees
We can now also expand the unitary operator $\bar{U}_F$ in powers of the perturbation,
\be
\bar{U}_F=\sum_{\nu=1}^\infty \bar{U}_F^{(\nu)}.
\ee
One finds
\bes
\bar{U}_F^{(0)}&=&1,
\\
\bar{U}_F^{(1)}&=&\bar{G}^{(1)},
\\
\bar{U}_F^{(2)}&=&\bar{G}^{(2)}+\frac{1}{2}\big[\bar{G}^{(1)}\big]^2,
\ees
where the second term of the last equation possesses matrix elements 
\be
\la\alpha'm'|\frac{1}{2}\big[\bar{G}^{(1)}\big]^2|\alpha m\rra
=\sum_{m''\ne m,m'}
    \frac{\la\alpha'|\Ho_{m'-m''}\Ho_{m''-m}|\alpha\ra}{(m'-m'')(m''-m)(\hbar\omega)^2},
\ee
which are finite also for $m'=m$.

The corresponding operators in $\mathcal{H}$ can be constructed by employing the relation
\be
\hat{A}(t) =\sum_m e^{im'\omega t} |\alpha'\ra\lla\alpha'm'|\bar{A}|\alpha 0\rra\la\alpha|
\ee
that is valid for operators $\bar{A}$ that are translationally invariant with respect to the ``photon'' number, 
$\lla\alpha'm+\Delta m|\bar{A}|\alpha m\rra = \lla\alpha\Delta m|\bar{A}|\alpha 0\rra$. 
In doing so, $\bar{U}_F$ and $\bar{G}$ translate into time periodic operators $\Uo_F(t)$ and
$\hat{G}(t)$ and Eq.~(\ref{eq:AnsatzUB}) into
\be
\Uo_F(t) \equiv \exp(\hat{G}(t))
\ee
(see also \ref{sec:operators}). The leading terms of the perturbation expansion take the form
\bes\label{eq:G(1)}
\hat{G}^{(1)}(t) &=& -\sum_{m\ne0}\frac{e^{im\omega t}}{m\hbar\omega}\Ho_m \, ,
\\
\hat{G}^{(2)}(t) &=& \sum_{m\ne0} \Bigg\{\frac{e^{im\omega t} \big[\Ho_0,\Ho_m\big]}{(m\hbar\omega)^2}
    +\frac{1}{2}\sum_{m'\ne 0,m} \frac{e^{i(m-m')\omega t}\big[\Ho_{-m'},\Ho_m\big]}{m(m-m')(\hbar\omega)^2}
        \Bigg\} 
\nonumber\\
\ees
and
\bes
\Uo_F^{(0)}(t)&=& 1,
\\
\Uo_F^{(1)}(t) &=&\hat{G}^{(1)}(t),
\\
\Uo_F^{(2)}(t) &=&\hat{G}^{(2)}(t) 
    +\frac{1}{2}\sum_{m\ne0}\sum_{m'\ne0}\frac{e^{i(m+m')\omega t}\Ho_{m'}\Ho_m}{m'm(\hbar\omega)^2}.
\ees
One can express these terms also as time integrals. For the leading order we obtain
\bes
\hat{G}^{(1)}(t)=\Uo_F^{(1)}(t)
&=& -\frac{1}{T}\int_{t_0}^{t_0+T}\!\rd t'\, \Ho(t') \sum_{m=1}^\infty\frac{2i\sin(m\omega (t-t'))}{m\hbar\omega}
\nonumber\\&=&
    - \frac{i\pi}{\hbar\omega} \frac{1}{T} \int_{t}^{t+T}\!\rd t'\, \Ho(t')
        \Big(1 - 2\frac{t-t'}{T}\Big).
\ees
In the final result we have separated a factor of $\frac{1}{T}$ representing the inverse 
integration time. It was obtained by setting the free parameter $t_0$ to $t_0=t$ allowing us to use
\be\label{eq:gr}
\sum_{k=1}^\infty \frac{\sin(kx)}{k}=\frac{\pi-x}{2} \qquad\text{for}\qquad 0<x<2\pi,
\ee
which is formula 1.441-1 of reference \cite{GradshteynRyzhik}.

One can now approximate $\Uo_F(t)$ up to a finite order $\tilde{\nu}$ by simply truncating the 
perturbative expansion of $\Uo_F(t)$ like $\Uo_F(t)\approx \sum_{\nu=0}^{\tilde{\nu}}\Uo_F^{(\nu)}(t)$. 
However, this approximation has the disadvantage that it does not preserve unitarity at any finite 
order $\tilde{\nu}$. In turn, truncating the expansion of $\hat{G}(t)$ leads to an approximation
\be\label{eq:Gapprox}
\Uo_F(t)\approx \exp\bigg(\sum_{\nu=1}^{\tilde{\nu}}\hat{G}^{(\nu)}\bigg) 
        \equiv \Uo^{[\tilde{\nu}]}_F(t) 
\ee
that gives rise to a unitary operator $\Uo^{[\tilde{\nu}]}_F(t)$ for every finite $\tilde{\nu}$.

The unitary two-point micromotion operator can be written like 
\be
\Uo_F(t,t')=\Uo_F(t)\Uo_F^\dag(t') \equiv \exp\big(\hat{F}(t,t')\big)
\ee
with anti-hermitian operator $\hat{F}(t,t')= -\hat{F}^\dag(t,t')$. Expanding $\hat{F}(t,t')$ in powers 
of the perturbation,
\be
\hat{F}(t,t')=\sum_{\nu=1}^\infty \hat{F}^{(\nu)}(t,t'),
\ee
and comparing the epxansion of $\exp\big(\hat{F}(t,t')\big)$ in powers of the perturbation with that of
$\exp\big(\hat{G}(t)\big)\exp\big(-\hat{G}(t')\big)$, 
one can identify
\bes
\hat{F}^{(1)}(t,t') &=& \hat{G}^{(1)}(t)-\hat{G}^{(1)}(t'),
\\
\hat{F}^{(2)}(t,t') &=& \hat{G}^{(2)}(t)-\hat{G}^{(2)}(t') 
                -\frac{1}{2}\big[\hat{G}^{(1)}(t),\hat{G}^{(1)}(t')\big],
\ees
and so on. This gives the explicit expressions for the leading orders
\bes\label{eq:MMO1}
\hat{F}^{(1)}(t,t') &=&  
            -\sum_{m\ne0}\frac{1}{m\hbar\omega}\bigg(e^{im\omega t}-e^{im\omega t'}\bigg)\Ho_m ,
\\\label{eq:MMO2}
\hat{F}^{(2)}(t,t') &=& 
    \sum_{m\ne0} \Bigg\{
        \frac{\Big(e^{im\omega t}-e^{im\omega t'}\Big)\big[\Ho_0,\Ho_m\big]}{(m\hbar\omega)^2}
\nonumber\\&&+\, 
      \frac{1}{2}\sum_{m'\ne 0,m}
    \frac{\Big(e^{i(m-m')\omega t}-e^{i(m-m') \omega t'}\Big)\big[\Ho_{-m'},\Ho_m\big]}
                        {m(m-m')(\hbar\omega)^2} \Bigg\}
\nonumber\\&&-\,\frac{1}{2}
                \sum_{m\ne0}\sum_{m'\ne0}
                \frac{e^{im\omega t-im'\omega t'}\big[\Ho_{-m'},\Ho_m\big]}{mm'(\hbar\omega)^2}.
\ees
An approximation preserving the unitarity of the micromotion operator reads
\be
\Uo_F(t,t')\approx \exp\bigg(\sum_{\nu=1}^{\tilde{\nu}}\hat{F}^{(\nu)}(t,t')\bigg) 
        \equiv \Uo^{[\tilde{\nu}]}_F(t,t').
\ee

\subsection{Effective Hamiltonian}
In order to obtain the effective Floquet Hamiltonian from Eq.~(\ref{eq:BloDi}), we need to compute 
the matrix elements (\ref{eq:HFloquet}) for $m=m'=0$, 
\be
H^F_{\alpha'\alpha}\equiv \la\alpha'|\Ho_F|\alpha\ra
=\lla \alpha'0|\bar{U}_F^\dag\bar{Q}\bar{U}_F|\alpha 0\rra = Q_{0,\alpha'\alpha}.
\ee
Expanding these matrix elements in powers of the perturbation, the leading terms 
$Q_{0,\alpha'\alpha}^{(\nu)}$ are given by Eqs.~(\ref{eq:Qeff0}), (\ref{eq:Qeff1}),
(\ref{eq:Qeff2}), and (\ref{eq:Qeff3}) of \ref{sec:perturbation}. Evaluating these expressions for the 
unperturbed problem (\ref{eq:unperturbed}), we obtain the perturbative expansion for the effective 
Hamiltonian
\be
\Ho_F = \sum_{\nu=0}^\infty \Ho^{(\nu)}_F,
\ee
with $\Ho^{(\nu)}_F=\sum_{\alpha'\alpha}|\alpha'\ra Q_{0,\alpha'\alpha}^{(\nu)}\la\alpha|$.
The leading terms are given by
\bes\label{eq:HF0}
\Ho_F^{(0)} &=& 0,
\\\label{eq:HF1}
\Ho_F^{(1)} &=& \Ho_0,
\\\label{eq:HF2}
\Ho_F^{(2)} &=& \sum_{m\ne0}\frac{\Ho_m\Ho_{-m}}{m\hbar\omega},
\\\label{eq:HF3}
\Ho_F^{(3)} &=& \sum_{m\ne0} \Bigg( \frac{\big[\Ho_{-m},\big[\Ho_0,\Ho_{m}\big]\big]}{2(m\hbar\omega)^2}
        +\sum_{m'\ne0,m}  
            \frac{\big[\Ho_{-m'},\big[\Ho_{m'-m},\Ho_m\big]\big]}{3mm'(\hbar\omega)^2} 
            \Bigg).
\nonumber\\
\ees
One can express these terms also in terms of time integrals. The leading order is given by the
time-averaged Hamiltonian,
\be\label{eq:HF1i}
\Ho_F^{(1)} = \frac{1}{T}\int_0^T \!\rd t\, \Ho(t).
\ee
The first correction takes the form
\bes\label{eq:HF2i}
\Ho_F^{(2)} &=& \frac{1}{T^2}\int_0^T\!\rd t_1\int_0^T\!\rd t_2\,
            \sum_{m\ne0}\frac{e^{-im\omega(t_1-t_2)} }{m\hbar\omega} 
                \Ho(t_1)\Ho(t_2)
\nonumber\\&=&
        \frac{1}{T^2}\int_0^T\!\rd t_1\int_0^{t_1}\!\rd t_2\, 
            \sum_{m\ne0}\frac{e^{-im\omega(t_1-t_2)} }{m\hbar\omega} 
                \big[\Ho(t_1),\Ho(t_2)\big]
\nonumber\\&=&\frac{2\pi}{i\hbar\omega}
        \frac{1}{2T^2}\int_0^T\!\rd t_1\int_0^{t_1}\!\rd t_2\, 
            \Big(1-2\frac{t_1-t_2}{T}\Big) 
            \big[\Ho(t_1),\Ho(t_2)\big],
\ees
where the sum over $m$ has been evaluated using Eq.~(\ref{eq:gr}) and where we have separated a factor
of $1/(2T^2)$ representing the inverse integration area. 
In $\tilde{\nu}$th order the effective Hamiltonian is approximated by
\be\label{eq:HFapprox}
\Ho_F \approx \sum_{\nu=0}^{\tilde{\nu}} \Ho^{(\nu)}_F \equiv \Ho^{[\tilde{\nu}]}_F.
\ee

The results obtained here via degenerate perturbation theory in the extended Floquet Hilbert space 
are equivalent to the high-frequency expansion derived in references \cite{RahavEtAl03,GoldmanDalibard14,
ItinKatsnelson14} by different means\footnote{There is a slight discrepancy, however, concerning the third-order 
correction to the effective Hamiltonian. The second term of our expression (\ref{eq:HF3}) is different from the 
corresponding term in equation (C.10) of reference \cite{GoldmanDalibard14}, where a spurious factor of 2 is found.}.

\subsection{Role of the driving phase}

An important property of the approximation (\ref{eq:HFapprox}) to the effective Hamiltonian is that it is independent of 
the driving phase. Namely, a shift in time
\be\label{eq:timeshift}
\Ho(t)\to\Ho'(t)=\Ho(t-t'),
\ee
which leads to 
\be
\Ho_{m}\to\Ho'_{m}=e^{-i m\omega t'}\Ho_{m},
\ee
does not alter the perturbation expansion of $\Ho_F$, 
\be
\Ho_F^{(\nu)}\to\Ho_F^{'(\nu)}=\Ho_F^{(\nu)}.
\ee
This is ensured by the structure of the perturbation theory, which restricts the products
$\Ho_{m_1}\Ho_{m_2}\cdots\Ho_{m_\nu}$ that contribute to $\Ho_F^{(\nu)}$ to those with
$m_1+m_2+\cdots m_\nu=0$. As an immediate consequence, also the approximate quasienergy spectrum, 
obtained from the diagonalization of $\Ho_F$, does not acquire a spurious dependence on the driving 
phase. In this respect, the high-frequency approximation obtained by truncating the high-frequency expansion of $\Ho_F$
at finite order, Eqs.~(\ref{eq:Gapprox}) and (\ref{eq:HFapprox}), is consistent with Floquet theory. 

A time shift does, however, modify the terms of the unitary operator $\Uo_F(t)$ in the expected way,
\be
\Uo^{(\nu)}_F(t)\to\Uo^{'(\nu)}_F(t)=\Uo^{(\nu)}_F(t-t'),
\ee
since
\be
\hat{G}^{(\nu)}(t)\to\hat{G}^{'(\nu)}(t)=\hat{G}^{(\nu)}(t-t').
\ee

\subsection{\label{sec:SpecMod}Quasienergy spectrum and Floquet modes}
From the approximate Floquet Hamiltonian one can now compute the quasienergy spectrum and the Floquet 
modes by solving the eigenvalue problem 
\be
\Ho_F^{[{\nu}]} |\tilde{u}_{n}\ra^{[{\nu}]} 
    = \varepsilon^{[{\nu}]}|\tilde{u}_n\ra^{[{\nu}]}.
\ee
One obtains
\be
\varepsilon_n\approx \varepsilon_n^{[{\nu}]}
\ee
and 
\be
|u_n(t)\ra \approx  \Uo_F^{[{\nu}']}|u_n(t)\ra^{[\nu]} 
    \equiv |u_n^{[{\nu}',{\nu}]}(t)\ra.
\ee
Here we have allowed that the order ${\nu}'$ of the approximate unitary operator $\Uo_F^{[\nu']}(t)$ 
describing the micromotion can be different from the order ${\nu}$ of the approximate Floquet 
Hamiltonian $\Ho_F^{[\nu]}$, which determines the Floquet spectrum and the dynamics on longer times. 
This corresponds to the approximation
\be\label{eq:HF2nu}
\Ho^F_{t_0}\approx 
    \Uo_F^{[{\nu}']}(t_0)\Ho_F^{[{\nu}]}\Uo_F^{[{\nu}']\dag}(t_0)
        \equiv \Ho^{F[{\nu}',{\nu}]}_{t_0}.
\ee
to the Floquet Hamiltonian $\Ho^F_{t_0}$. 

The reason why it is generally useful to choose $\nu'$ independent of $\nu$ is the following. In
high-frequency approximation the time evolution from $t_0$ to $t$ is described by
\be
|\psi(t)\ra\approx 
    \sum_n  c_n^{[\nu',\nu]}\,
        |u_n^{[\nu',\nu]}(t)\ra  \, e^{-i\varepsilon_n^{[\nu]} (t-t_0)/\hbar},
\ee
with $c_n^{[\nu',\nu]} = \la u_n^{[\nu',\nu]}(t_0)|\psi(t_0)\ra$. The accuracy with which the
expression $c_n^{[\nu',\nu]}\,|u_n^{[\nu',\nu]}(t)\ra$ captures the true micromotion of the system does 
not depend on the time span $(t-t_0)$ of the integration, simply because this expression is time periodic.
In turn, with increasing integration time $(t-t_0)$, the approximate phase factors
$e^{-i\varepsilon_n^{[\nu]} (t-t_0)/\hbar}$ will deviate more and more from their actual value
$e^{-i\varepsilon_n (t-t_0)/\hbar}$. Thus, the longer the time span $t-t_0$ the better should 
be the approximation $\varepsilon_n\approx\varepsilon_n^{[\nu]}$, that is the larger should be $\nu$. In 
contrast, the order $\nu'$ can be chosen independently of $(t-t_0)$.

\section{\label{sec:FM}Relation to the Floquet-Magnus expansion}

In this section we relate the high-frequency expansion of $\Ho_F$ and $\hat{G}(t)$ to the Floquet-Magnus expansion
\cite{CasasEtAl00} (see also \cite{Feldman84,BlanesEtAl09,BukovEtAl14}).  A discussion of this issue can also be 
found in references \cite{RahavEtAl03,GoldmanDalibard14,GoldmanEtAl15}. Recently, the Floquet-Magnus expansion 
has been employed frequently for the treatment of quantum Floquet systems. The starting point 
of the Floquet-Magnus expansion is the form (\ref{eq:Evolution}) of the time evolution operator,
\bes\label{eq:Umagnus}
\Uo(t,t_0)&=& \Uo_F(t,t_0)\,\exp\bigg(-\frac{i}{\hbar}(t-t_0) \Ho^F_{t_0 }\bigg)
\nonumber\\
&=& \exp\Big(\hat{F}(t,t_0)\Big)\,\exp\bigg(-\frac{i}{\hbar}(t-t_0) \Ho^F_{t_0 }\bigg).
\ees
Then both $\hat{F}(t,t_0)$ and $\Ho^F_{t_0 }$ are expanded in powers of the Fourier transform of the Hamiltonian. 
Note that in references \cite{CasasEtAl00, BlanesEtAl09} the notation $P(t)=\Uo_F(t,0)$,
$\Lambda(t)=\hat{F}(t,0)$, and $F=-\frac{i}{\hbar}\Ho^F_{0}$ is used, implicitly assuming $t_0=0$. 

The Floquet-Magnus expansion of $\hat{F}(t,t_0)$ is reproduced by our expressions (\ref{eq:MMO1}) and
(\ref{eq:MMO2}). The Floquet-Magnus expansion of $\Ho^F_{t_0}$ can also be 
obtained within our formalism. Namely, expanding $\Ho^F_{t_0}$ in powers of the perturbation
\be\label{eq:FMH}
\Ho^F_{t_0} = \sum_{\nu=1}^\infty \Ho^{F(\nu)}_{t_0},
\ee
gives
\bes
\Ho^{F(1)}_{t_0}&=&\Ho_F^{(1)},
\\ \label{eq:HFM2exp}
\Ho^{F(2)}_{t_0}&=&\Ho_F^{(2)}+\Uo_F^{(1)}(t_0)\Ho^{(1)}_F+\Ho^{(1)}_F\Uo_F^{(1)\dag}(t_0),
\\
\Ho^{F(3)}_{t_0}&=&\Ho_F^{(3)}+\Uo_F^{(2)}(t_0)\Ho_F^{(1)}+\Ho_F^{(1)}\Uo_F^{(2)\dag}(t_0)
\nonumber\\&&
        +\,\Uo_F^{(1)}(t_0)\Ho_F^{(1)}\Uo_F^{(1)\dag}(t_0),
\nonumber\\
\ees
and so on. From these expressions one obtains 
\bes\label{eq:HFM1}
\Ho^{F(1)}_{t_0}&=&\Ho_0,
\\ \label{eq:HFM2}
\Ho^{F(2)}_{t_0}&=& \sum_{m\ne0}\frac{1}{m\hbar\omega} 
            \bigg(\Ho_m\Ho_{-m} + e^{im\omega t_0} \big[\Ho_0,\Ho_m\big]\bigg)
\ees
and, in a subsequent step, also
\bes
\Ho^{F(1)}_{t_0}&=& \frac{1}{T}\int_{t_0}^{t_0+T}\!\rd t_1 \Ho(t_1),
\\ 
\Ho^{F(2)}_{t_0}&=& \frac{2\pi}{i\hbar\omega} \frac{1}{2T^2}
            \int_{t_0}^{t_0+T}\!\rd t_1\int_{t_0}^{t_0+t_1}\!\rd t_2\, 
             \big[\Ho(t_1),\Ho(t_2)\big],   
\ees
where we have again employed Eq.~(\ref{eq:gr}). For $t_0=0$ these expressions correspond to those of 
references \cite{CasasEtAl00, BlanesEtAl09}. 

Truncating the Floquet-Magnus expansion after the finite order $\tilde{\nu}$, the Floquet Hamiltonian is 
approximated like
\be\label{eq:FMapprox}
\Ho^F_{t_0}\approx\sum_{\nu=1}^{\tilde{\nu}} \Ho^{F(\nu)}_{t_0} 
        \equiv \Ho^{FM[\tilde{\nu}]}_{t_0}.
\ee
However, even though it is derived from a systematic expansion, this approximation is plagued by the 
following problem. For any finite order $\tilde{\nu}\ge2$, the spectrum of the approximate Floquet 
Hamiltonian $\Ho^{FM[\tilde{\nu}]}_{t_0}$ possesses an artifactual dependence on $t_0$, or equivalently on 
the driving phase. This is not consistent with the spectrum of the exact Floquet Hamiltonian $\Ho_{t_0}^F$, which is 
independent of the driving phase. In second order, the $t_0$ dependence enters with the second term of Eq.~(\ref{eq:HFM2}).
Let us consider, for example, a periodic Hamiltonian with even time dependence, $\Ho(t)=\Ho(-t)$, so that
$\Ho_m=\Ho_{-m}$. In this case $\Ho^{F(2)}_{t_0}$ vanishes for $t_0$ being an integer multiple of $\pi/\omega$,
while it is generally finite for other values of $t_0$. Therefore, generally the Floquet Hamiltonians
$\Ho^{M[\tilde{\nu}]}_{Ft_0}$ and $\Ho^{M[\tilde{\nu}]}_{Ft_0'}$ obtained from the Floquet-Magnus 
approximation for times $t_0\ne t_0'$ are not related to each other by a unitary transformation, as it is 
the case for the exact Floquet Hamiltonian. 

The origin of this spurious $t_0$ dependence lies in the fact that the
expansion~(\ref{eq:FMH}) of the Floquet Hamiltonian implies also an expansion
$\Uo_F(t)= 1+\Uo_F^{(1)}+\Uo_F^{(2)}(t)\cdots$ of the unitary operator $\Uo_F(t)$. At any finite order, such an expansion 
does not preserve unitarity and, thus, the spectrum of the approximate Floquet Hamiltonian $\Ho^{FM[\tilde{\nu}]}_{t_0}$ 
deviates from the $\tilde{\nu}$th-order spectrum obtained by diagonalizing the approximate effective Hamiltonian
$\Ho_F^{[\tilde{\nu}]}$ given by Eq.~(\ref{eq:HFapprox}). 

This observation can be traced back further to the ansatz~(\ref{eq:Umagnus}) for the time evolution operator. 
Bi-partitioning the time-evolution operator into two exponentials like in Eq.~(\ref{eq:Umagnus}) does not 
allow for disentangling the phase evolution from the micromotion. This is different for the tri-partitioning
ansatz \label{RahavEtAl03}
\bes\label{eq:tri}
\Uo(t,t_0)&=& \Uo_F(t)\,\exp\bigg(-\frac{i}{\hbar}(t-t_0) \Ho^F\bigg)\Uo_F^\dag(t_0)
\nonumber\\
&=& \exp\Big(\hat{G}(t)\Big)\,\exp\bigg(-\frac{i}{\hbar}(t-t_0) \Ho^F\bigg) \exp\Big(-\hat{G}(t_0)\Big),
\ees
which underlies the perturbative approach presented in the previous section. In the tri-partitioning 
ansatz~(\ref{eq:tri}), first $\Uo_F^\dag(t_0)$ transforms the state into a ``reference frame'' where by 
construction no micromotion is present. Then the phase evolution is generated by the effective Hamiltonian, 
before at time $t$ the state is finally rotated back to the original frame by $\Uo_F(t)$. 
In contrast $\Ho^F_{t_0}$, as it appears in the ansatz (\ref{eq:Umagnus}), carries also information about the
micromotion. This fact is somewhat hidden, when the $t_0$ dependence of the Floquet Hamiltonian is not
written out explicitly like in Ref.~\cite{BlanesEtAl09}, where $t_0=0$ is assumed.

However, since we know that the effective Hamiltonian $\Ho_F$ and the Floquet Hamiltonian $\Ho^F_{t_0}$ 
possess the same spectrum, we also know that, when expanding both $\Ho_F$ and $\Ho^F_{t_0}$ in powers of
the inverse frequency, also the spectra will coincide up to this order. This means that the $t_0$-dependent 
second term of Eq.~(\ref{eq:HFM2}) will not cause changes of the spectrum within the second order
($\propto\omega^{-1}$). Instead this second term can contribute to the third-order correction of the 
quasienergy spectrum, together with the terms of $\Ho_{t_0}^{(3)}$. This argument generalizes to higher 
orders. 

Let us illustrate our reasoning using a simple example. A spin-1/2 system shall be described by the
time-periodic Hamiltonian
\be
\Ho(t)= a \hat{S_x} + b \cos(\omega t) \hat{S}_y, 
\ee
with spin operators $\hat{S}_i$ and Fourier components
\be
\Ho_0 = a \hat{S_x},
\qquad
\Ho_1 = \Ho_{-1} =  \frac{b}{2} \hat{S}_y ,
\qquad
\Ho_{m}=0 \;\;\text{for}\;\; |m|\ge2.
\ee
According to equations (\ref{eq:HF1}) and (\ref{eq:HF2}), in second order
the effective Hamiltonian is approximated by 
\be
\Ho_F\approx\Ho_F^{[2]} =  \Ho_F^{(1)}+ \Ho_F^{(2)} ,
\ee
with
\be
\Ho_F^{(1)} = a \hat{S}_x, \qquad \Ho_F^{(2)} = 0.
\ee
Here the second-order term $\Ho_F^{(2)}$ vanishes since $[\Ho_m,\Ho_{-m}]=0$. In second order the 
quasienergy spectrum is, thus, approximated by
\be
\varepsilon_\pm\approx\varepsilon^{[2]}_\pm=\pm \frac{1}{2} \hbar a .
\ee
In contrast, the second-order approximation of the Floquet Hamiltonian based on the Floquet-Magnus expansion,
\be
\Ho^F_{t_0}\approx\Ho_{t_0}^{FM[2]} =  \Ho_{t_0}^{F(1)} + \Ho_{t_0}^{F(2)}, 
\ee
does contain a second-order term. Namely, from Eqs.~(\ref{eq:HFM1}) and (\ref{eq:HFM2}) one obtains
\be
\Ho_{t_0}^{F(1)} = a \hat{S}_x, \qquad \Ho_{t_0}^{F(2)} = -\frac{ab}{\omega} \sin(\omega t_0) \hat{S}_z.
\ee
It leads to the approximation of the quasienergy spectrum 
\bes
\varepsilon_\pm\approx\varepsilon^{FM[2]}_\pm &=& \pm\frac{1}{2} \hbar a \sqrt{1 + \big[(b/\omega) \sin(\omega t_0)\big]^2}
\nonumber\\
&=&  \pm\frac{1}{2} \hbar a \Big(1 + \frac{1}{2}\big[(b/\omega)\sin(\omega t_0)\big]^2 + \cdots \Big).
\ees
We can now make several observations that illustrate the reasoning of the previous paragraphs. First, we 
can see that $\varepsilon^{FM[2]}_\pm$ coincides with $\varepsilon^{[2]}_\pm$ within the order of the 
approximation. Deviations that occur are proportional to $\omega^{-2}$, while our second-order approximation 
should provide the correct terms up to the power $\omega^{-1}$. Second, despite the presence of a finite 
second-order term $\Ho_{t_0}^{F(2)}$ proportional to $\omega^{-1}$, $\varepsilon^{FM[2]}_\pm$ does not 
contain a correction $\propto\omega^{-1}$. This is consistent with the fact that $\Ho_F^{[2]}$ and, as a 
consequence, also $\varepsilon^{[2]}_\pm$ do not contain a second-order term. Third, we can see that, unlike 
the exact quasienergy spectrum, the approximate spectrum $\varepsilon^{FM[2]}_\pm$ depends on the time $t_0$ 
and, thus, also on the driving phase. However, this dependence on $t_0$ (or the driving phase) occurs only in terms
$\propto\omega^{-2}$ that are not reproduced correctly within the second-order approximation. In a third-order 
approximation, the spectrum will be captured correctly and be independent of the driving phase up to the power
$\omega^{-2}$ and so on. 

As a further example, we will discuss the circularly driven hexagonal lattice in the next section. There, we will see that 
the spurious driving-phase dependence of the Floquet-Magnus expansion will, additionally, also induce a spurious breaking 
of the rotational symmetry of the quasienergy dispersion relation (Section \ref{sec:HexFM}). Thus, even a weak $t_0$ 
dependence can seemingly change the properties of the system in a fundamental way. Therefore, the Floquet-Magnus 
approximation should be used with care. The high-frequency approximation derived in the previous section \ref{sec:HF} 
does not suffer from this problem.

\section{\label{sec:example}Example: Circularly driven hexagonal lattice}

\begin{figure}[t]
\includegraphics[width=1\linewidth]{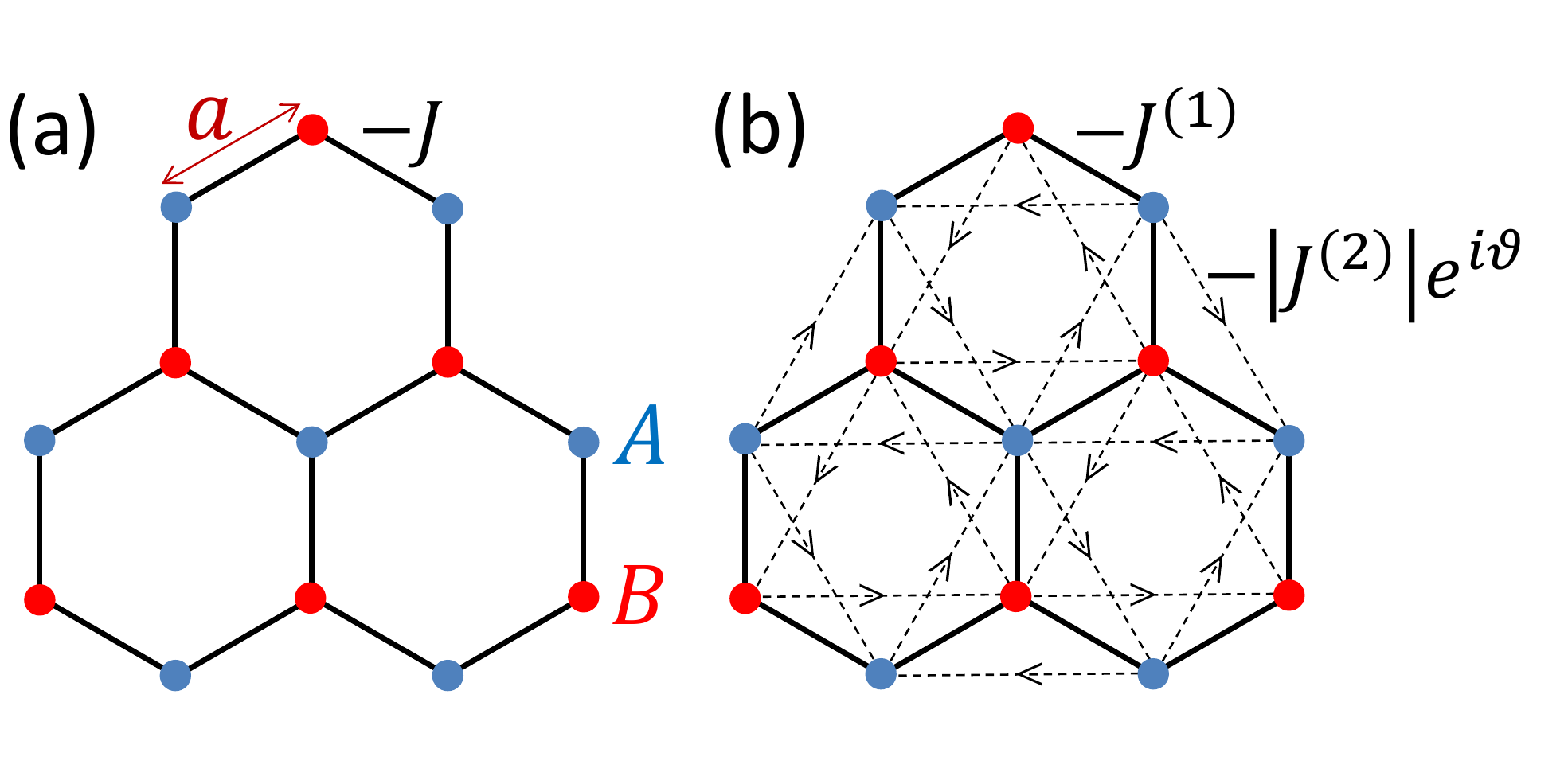}
\centering
\caption{\label{fig:hexagonal} Hexagonal lattice with sublattice A (blue) and B (red). (a) The 
Hamiltonian $\Ho_\text{kin}$ of the undriven model possesses real tunneling matrix elements $-J$ between 
neighboring sites. (b) The effective Hamiltonian $\Ho_F\approx\Ho_F^{(1)}+\Ho^{(2)}_F$ of the driven 
system features modified real tunneling matrix elements $-J^{(1)}$ between nearest neighbors, 
contained in $\Ho_F^{(1)}$, and complex tunneling matrix elements $-|J^{(2)}|e^{i\theta}$ (or 
$-|J^{(2)}|e^{-i\theta}$), contained in $\Ho_F^{(2)}$, for tunneling in anticlockwise (or clockwise) 
direction around the hexagonal plaquette.}
\end{figure}

In this section, we will discuss an instructive example of the physics of particles hopping on 
a hexagonal lattice [see Fig.~\ref{fig:hexagonal}(a)] subjected to a circular time-periodic force 
\be
  \bF(t) = -F[\cos(\omega t){\bm e}_x+\sin(\omega t){\bm e}_y].
\ee
For a system of charged electrons such a force can be realized by applying circularly 
polarized light, whereas for a system of neutral particles (atoms in an optical lattice or photons 
in a wave guide) it can be achieved as an inertial force via circular lattice shaking \cite{EckardtEtAl10,StruckEtAl11,
RechtsmanEtAl13,JotzuEtAl14}. The driven hexagonal lattice is particularly interesting, as it is the prototype of a 
Floquet topological insulator \cite{OkaAoki09,KitagawaEtAl10,KitagawaEtAl11,LindnerEtAl11,CayssolEtAl13}. It was pointed 
out by Oka and Aoki \cite{OkaAoki09} that a non-vanishing forcing strength $F$ opens a topological gap in the band 
structure of the effective Hamiltonian. As a consequence, the system possesses a quantized Hall conductivity, when the 
lowest band is filled completely with fermions. While the original proposal \cite{OkaAoki09} is considering graphene 
irradiated by circularly polarized light (see also Ref.~\cite{PerezPiskunowEtAl14}), the topologically non-trivial band 
structure described by the effective Hamiltonian has been probed experimentally in other systems: with classical light in 
a hexagonal lattice of wave guides \cite{RechtsmanEtAl13} and with ultracold fermionic atoms in a circularly shaken 
optical lattice \cite{JotzuEtAl14}. 

We have decided to discuss the circularly driven hexagonal lattice here, even though its single-particle 
physics has been described in detail already elsewhere \cite{OkaAoki09,KitagawaEtAl11,ZhengZhai14,JotzuEtAl14}, because 
of several reasons. First, it is a paradigmatic example of a system where the second-order high-frequency correction to 
the effective Hamiltonian gives rise to qualitatively new physics. Second, since both directions, $x$ and $y$, 
are driven with a phase lag of $\pi/2$, the model is suitable to illustrate the difference between the 
high-frequency expansion advertised here and the Floquet-Magnus expansion. And third, it allows us 
to set the stage for the ensuing discussion (see Sect.~\ref{sec:interactions}) of 
the role of interactions, which have been discussed controversially recently
\cite{GrushinEtAl14,DAlessio14}. This issue includes two aspects: the impact of interactions on 
the validity of the high-frequency expansion as well as how interactions appear in the 
high-frequency expansion.

Let us consider the driven tight-binding Hamiltonian 
\be
\Ho_\text{dr}(t) = -\sum_{\la\ell'\ell\ra}  J\aa_{\ell'}\ao_\ell +\sum_\ell v_\ell(t)\no_\ell.
\ee
The first term describes the tunneling kinetics, with the sum running over all directed links 
$\la\ell'\ell\ra$ connecting a site $\ell$ to its nearest neighbor $\ell'$ 
on the hexagonal lattice depicted in Fig.~\ref{fig:hexagonal}(a). 
Here $\ao_\ell$ is the annihilation operator for a particle (boson or fermion) at the lattice 
site $\ell$ located 
at $\br_\ell$, and the tunneling parameter $J$ is real and positive. The second sum runs over the
lattice sites and describes the effect of the driving 
force in terms of the time-periodic on-site potential $v_\ell(t)=-\br_\ell\cdot\bF(t)$ and
the number operator $\no_\ell=\aa_\ell\ao_\ell$. The direction of the vector pointing from site $\ell$
to a neighbor $\ell'$ defines an angle $\varphi_{\ell'\ell}$,
\be
\br_{\ell'}-\br_\ell \equiv a [\cos(\varphi_{\ell'\ell}){\bm e}_x+\sin(\varphi_{\ell'\ell}){\bm e}_y],
\ee  
with $\varphi_{\ell\ell'}=\varphi_{\ell'\ell}+\pi$.
This angle determines the temporal driving phase of the relative potential modulation between both sites,
\be
v_{\ell'}(t)-v_\ell(t) = Fa\cos(\omega t -\varphi_{\ell'\ell}).
\ee

\subsection{Change of gauge}

As will be seen shortly, we are interested in the regime of strong forcing, where the amplitude
$K\equiv Fa$ of the relative potential modulation between two 
neighboring sites is comparable to or larger than $\hbar\omega$. Therefore, the Hamiltonian $\Ho_\text{dr}(t)$ 
is not a suitable starting point for the high-frequency approximation. 

A remedy is provided by 
a gauge transformation with the time-periodic unitary operator \cite{EckardtEtAl10}
\be\label{eq:Ugauge}
\Uo(t) = \exp\Big(i \sum_\ell \chi_\ell(t) \no_\ell \Big),
\ee
where
\bes
\chi_\ell(t)  &=& -\int_0^t \!\rd t'\,\frac{v_\ell(t')}{\hbar}
         + \frac{1}{T}\int_0^T\!\rd t'' \int_0^{t''} \!\rd t'\,\frac{v_\ell(t')}{\hbar}
\nonumber\\     
    &=&  \frac{F\br_\ell}{\hbar\omega}\cdot[- \sin(\omega t){\bm e}_x + \cos(\omega t){\bm e}_y].
\ees
This gauge transformation induces a time-dependent shift in quasimomentum, and the second integral 
has been included to eliminate an overall quasimomentum drift. It provides a constant that subtracts the
zero-frequency component of the first integral, thus making the time average of $\chi_\ell(t)$ over 
one driving 
period vanish. One arrives 
at the translationally invariant time-periodic Hamiltonian
\be\label{eq:gauge}
\Ho(t)=
\Uo^\dag(t)\Ho_\text{dr}(t)\Uo(t) -{ i\hbar\,} \Uo^\dag(t)\dot{\Uo}(t)
=
-\sum_{\la\ell'\ell\ra} J e^{i\theta_{\ell'\ell}(t)}\aa_{\ell'}\ao_\ell .
\ee
Here the scalar potential $v_\ell(t)$ is absent while the driving force is captured by 
the time-periodic Peierls phases 
\be
\theta_{\ell'\ell}(t) = \chi_\ell(t)-\chi_{\ell'}(t)
    =\frac{K}{\hbar\omega} \sin(\omega t-\varphi_{\ell'\ell}).
\ee
Now we are in the position to apply the high-frequency approximation, even for $K\gg\hbar\omega$.
The actual requirement is that $\hbar\omega$ must be large compared to the tunneling matrix 
element $J$, which determines both the spectral width of $\Ho_0$ and the strength of the 
coupling terms $\Ho_m$ with $m\ne0$.

\subsection{\label{sec:hexHeff}Effective Hamiltonian}

\begin{figure}[t]
\includegraphics[width=0.6\linewidth]{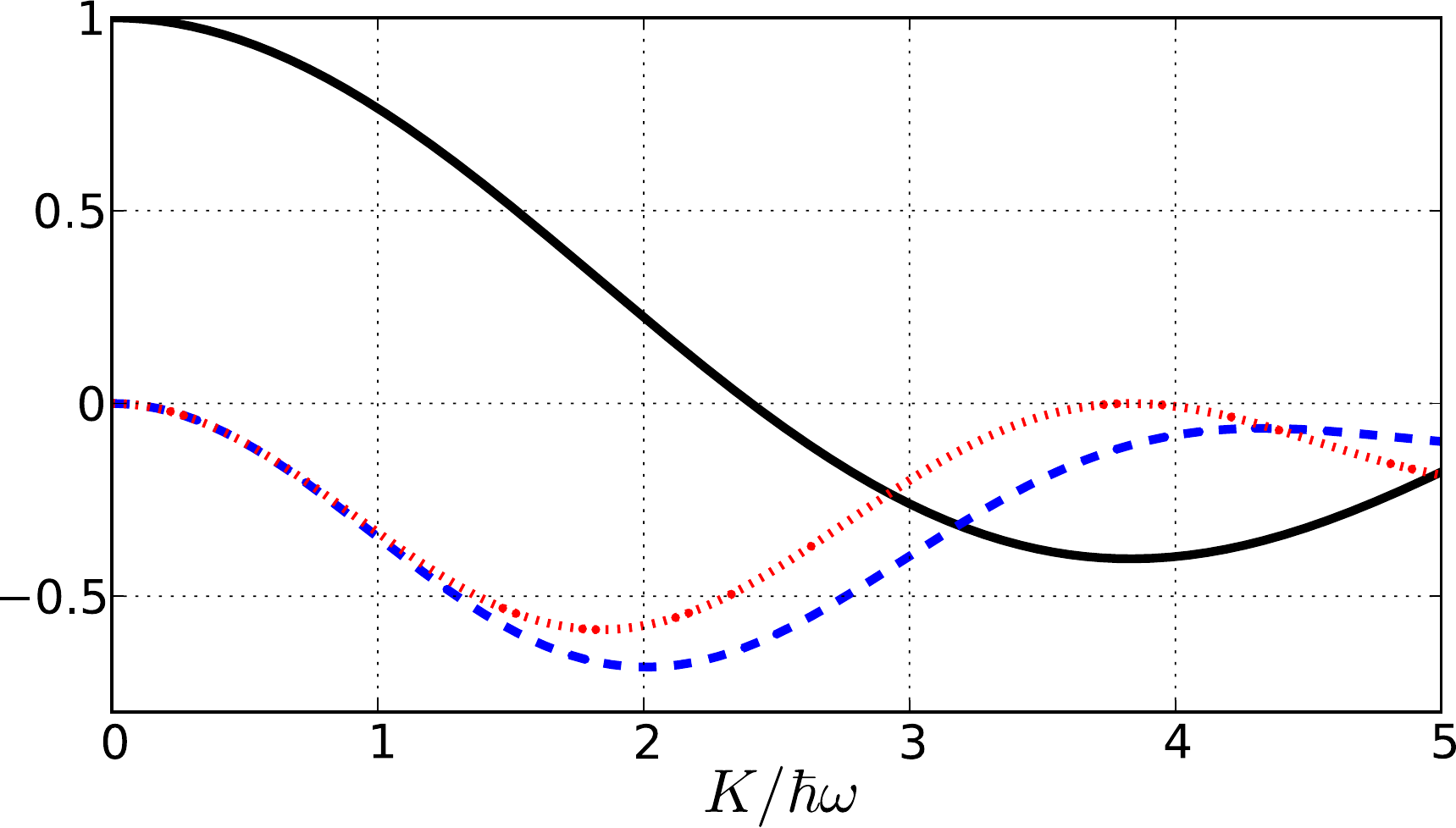}
\centering
\caption{\label{fig:Jeff} Effective tunneling matrix elements $J^{(1)}/J$ (solid black line) 
and $J^{(2)}/ (J^2/\hbar\omega)$ (dashed blue line), as well as the leading term
$-\sqrt{3}\bJ_1^2(K/\hbar\omega)$ contributing to $J^{(2)}/ (J^2/\hbar\omega)$ (dotted red
 line).}
\end{figure}

The leading term in the expansion of the effective Hamiltonian is according to Eq.~(\ref{eq:HF1})
given by the time-average of the driven Hamiltonian 
\be\label{eq:H1hex}
\Ho_F^{(1)} =\Ho_0= -\sum_{\la\ell'\ell\ra} J^{(1)}\aa_{\ell'}\ao_\ell.
\ee
It corresponds to the undriven Hamiltonian with a modified effective tunneling matrix element
\be\label{eq:Jeff1}
J^{(1)} = J \bJ_0\big({\T\frac{K}{\hbar\omega}}\big),
\ee
where $\bJ_n$ denotes a Bessel function of integer order $n$. This result was obtained by employing the relation
\be
\exp(ir\sin(s))=\sum_{k=-\infty}^\infty \bJ_k(r)\exp(iks).
\ee
This Bessel-function-type renormalization of the tunnel matrix element, see Fig.~\ref{fig:Jeff} for a 
plot, allows to effectively reduce or even ``switch off'' completely the nearest neighbor tunneling 
matrix element. This effect is known as dynamic localization \cite{DunlapKenkre86}, coherent destruction of 
tunneling \cite{GrossmannEtAl91,GrifoniHaenggi98}, or band collapse \cite{Holthaus92}. It has been observed in the
coherent expansion of a localized Bose condensate in a shaken optical lattice \cite{LignierEtAl07}. The
effect has also been used to induce the transition between a bosonic superfluid to a Mott insulator (and
back) by shaking an optical lattice \cite{EckardtEtAl05b,ZenesiniEtAl09}. The possibility to make the 
tunneling matrix element negative has moreover been exploited to achieve kinetic frustration in a 
circularly forced triangular lattice and to mimic antiferromagnetism with spinless bosons 
\cite{EckardtEtAl10,StruckEtAl11}.

The second-order contribution to the effective Hamiltonian is given by Eq.~(\ref{eq:HF2}) and can be 
written like
\be
  \Ho_F^{(2)}= \sum_{m=1}^\infty \frac{1}{m\hbar\omega}\big[\Ho_m,\Ho_{-m}\big],
\ee
with the Fourier components of the Hamiltonian reading
\be\label{eq:Hm}
\Ho_m=-\sum_{\la\ell'\ell\ra} J \bJ_m\big({\T\frac{K}{\hbar\omega}}\big)e^{-im\varphi_{\ell'\ell}} 
            \,\aa_{\ell'}\ao_\ell.
\ee
By using the relation $\big[\aa_k\ao_l,\aa_m\ao_n\big]=\delta_{lm}\aa_k\ao_n-\delta_{kn}\aa_m\ao_l$, 
which holds both for bosonic and fermionic operators $\ao_\ell$, as well as $\bJ_{-m}(x)=(-)^m\bJ_m(x)$, one 
arrives at 
\be\label{eq:H2hex}
\Ho^{(2)}_F=-\sum_{\lla\ell'\ell\rra} J^{(2)}_\text{\lla\ell'\ell\rra}\, \aa_{\ell'}\ao_\ell,
\ee
where the sum runs over next-nearest neighbors $\ell'$ and $\ell$. The effective tunneling matrix 
element is given by 
\be
    J_{\lla\ell'\ell\rra}^{(2)} = \frac{J^2}{\hbar\omega}\sum_{m=1}^\infty
    \frac{1}{m}\bJ_m^2\big({\T\frac{K}{\hbar\omega}}\big) 
    2i\sin\big( m[\varphi_{\ell'k}-\varphi_{\ell k}]\big),
\ee
where $k$ denotes the intermediate lattice site between $\ell'$ and $\ell$, via which the 
second-order tunneling process occurs\footnote{In other lattice geometries several two-step paths 
between $\ell'$ and $\ell$ can exist, in this case one has to sum over all of them.}. One can immediately 
see that the tunneling matrix elements $ J_{\lla\ell'\ell\rra}^{(2)}$ are purely imaginary and that they 
depend, as an odd function, on the relative angle $ \varphi_{\ell'k}-\varphi_{\ell k}$ only. This 
relative angle is given by $ \varphi_{\ell'k}-\varphi_{\ell k}= - \sigma_{\ell'\ell}\frac{2\pi}{3}$, with sign 
$\sigma_{\ell'\ell}=+1$ ($\sigma_{\ell'\ell}=-1$) for tunneling in anticlockwise (clockwise) direction 
around a hexagonal lattice plaquette. Therefore, one finds 
\be
J_{\lla\ell'\ell\rra}^{(2)}= i J^{(2)}\sigma_{\ell'\ell}= |J^{(2)}|e^{i\sigma_{\ell'\ell}\theta}
\ee
forming the pattern of effective tunneling matrix elements depicted in Fig.~\ref{fig:hexagonal}(b). Here
\be\label{eq:Jeffprime}
J^{(2)}= - \frac{J^2}{\hbar\omega}\sum_{m=1}^\infty
        \frac{1}{m}\bJ_m^2\big({\T\frac{K}{\hbar\omega}}\big) 2\sin\big(m 2\pi/3\big)
        \simeq-\sqrt{3}\frac{J^2}{\hbar\omega}\bJ_1^2\big({\T\frac{K}{\hbar\omega}}\big)
\ee
and 
\be
\theta=\mathrm{sign}(J^{(2)})\frac{\pi}{2}.
\ee
Since $|\bJ_m(x)|$ decays like $|x|^{|m|}$ with respect to the order $m$, for sufficiently small
$K/\hbar\omega$ the sum is to good approximation exhausted by its first term, as is demonstrated
also in Fig.~\ref{fig:Jeff}.

The approximate effective Hamiltonian \cite{KitagawaEtAl11}
\be\label{eq:Heff[2]}
\Ho_F\approx \Ho_F^{(1)}+\Ho_F^{(2)},
\ee
as it is depicted in Fig.~\ref{fig:hexagonal}(b), directly corresponds to the famous Haldane model \cite{Haldane88} 
(see also reference \cite{Cayssol13}), being the prototype of a topological Chern insulator
\cite{HasanKaneEtAl10,QiZhang11}. The next-nearest neighbor tunneling matrix elements open a gap between the two
low-energy Bloch bands of the hexagonal
lattice, such that the bands acquire topologically non-trivial properties of a Landau level characterized by
non-zero integer Chern number \cite{ThoulessEtAl82}. As a consequence, the system features chiral edge 
states, as they have been observed experimentally with optical wave guides \cite{RechtsmanEtAl13}, and a 
finite Hall conductivity, as it has been observed with ultracold fermionic atoms \cite{JotzuEtAl14}, which 
is quantized for a completely filled lower band. The fact that circular forcing can induce such non-trivial 
properties to a hexagonal lattice has been pointed out in Ref.~\cite{OkaAoki09}. This is the first proposal 
for a Floquet-topological insulator \cite{CayssolEtAl13} (later proposals include
Refs.~\cite{KitagawaEtAl10,LindnerEtAl11}). These systems can be defined as driven lattice systems with the effective 
Hamiltonian featuring new matrix elements that open topologically non-trivial gaps in the Floquet-Bloch band structure. 
Such new matrix elements appear in second (or higher) order of the high-frequency approximation that capture processes 
where a particle tunnels twice (or several times) during one driving period and that are of the order of
$\sim J^2/\hbar\omega$. Therefore, Floquet topological insulators require the driving frequency to be at most
moderately larger than the tunneling matrix element $J$. This is different for another class of schemes for
the creation of artificial gauge fields and topological insulators recently pushed forward mainly in the
context of ultracold quantum gases \cite{AidelsburgerEtAl11,Kolovsky11,BermudezEtAl11,StruckEtAl12,HaukeEtAl12b,
AidelsburgerEtAl13,MiyakeEtAl13,AtalaEtAl14,AidelsburgerEtAl15}. In these schemes non-trivial effects enter already in 
the leading first-order term of the high-frequency expansion, such that they work also in the high-frequency limit
$\hbar\omega\gg J$. 

The non-interacting driven hexagonal lattice considered in this section can easily be solved numerically, without further 
approximation. The driven Hamiltonian (\ref{eq:gauge}) obeys discrete translational lattice, with two sublattice states 
per lattice cell. As a consequence, the single-particle state space is divided into 
uncoupled sectors of sharp quasimomentum, each containing two states. The problem is reduced to that of a family of 
driven two-level systems labeled by the quasimomentum wave vector $\bk$. The high-frequency approximation
(\ref{eq:Heff[2]}) is nevertheless useful. First, it gives rise to an analytical approximation to the Floquet 
Hamiltonian, directly corresponding to the paradigmatic Haldane model \cite{KitagawaEtAl11,JotzuEtAl14}. Second, as we 
will argue in the next paragraph, the second-order approximation captures already the essential physics of the
non-interacting system in the regime of large frequencies. And third, it allows for taking into account also some of the 
effects related to interactions (see section \ref{sec:IntCorr}).

Let us briefly argue why the approximate effective Hamiltonian (\ref{eq:Heff[2]}) captures the essential properties 
of the full one for the translationally invariant non-interacting system in the limit of large driving frequencies. The
two-level Hamiltonian acting in the single-particle space of states with quasimomentum $\bk$ is represented by a $2\times2$ matrix of the general form
$
h(\bk,t)=h_0(\bk,t) + \bm{h}(\bk,t)\cdot{\bm \sigma}
$,
with ${\bm\sigma}$ denoting the vector of Pauli matrices and vector $\bm{h}=(h_x,h_y,h_z)$. If the elements
$h_i(\bk,t)\sim J$ are small compared to $\hbar\omega$, the perturbation expansion can be expected to converge. 
It is then left to argue that corrections beyond the second-order approximation (\ref{eq:Heff[2]}) do not lead to 
qualitatively new behavior. This can be done employing the arguments by Haldane \cite{Haldane88}
(see also reference \cite{Cayssol13}). In the subspace of quasimomentum $\bk$, the effective Hamiltonian $\Ho_F$ is 
represented by a time-independent matrix 
\be
h_F(\bk)=h_{F0}(\bk) + \bm{h}_F(\bk)\cdot{\bm \sigma}.
\ee 
Its components determine the single-particle quasienergy dispersion relation 
\be\label{eq:DispF}
\varepsilon^F_\pm(\bk)=h_{F0}(\bk)\pm\sqrt{|{\bm h}_F(\bk)|},
\ee 
with the two bands labeled by $-$ and $+$.
The leading approximation of the effective Hamiltonian, $\Ho^{(1)}_F$, describes a hexagonal lattice 
with real nearest-neighbor tunneling matrix element $-J^{(1)}\sim J$ [see Fig.~\ref{fig:hexagonal}(b)]. It gives rise to 
a contribution ${\bm h}^{(1)}_F(\bk)$ to the vector ${\bm h}_F(\bk)$. Due to the fact that $\Ho^{(1)}_F$ obeys
space-inversion and time-reversal symmetry, two inequivalent quasimomenta $\bk_\pm$ (the corners $K$ and $K'$ of the 
first Brillouin zone) exist, where ${\bm h}^{(1)}(\bk_\pm)=0$ \cite{Cayssol13}. These are the Dirac points where the 
bands described by ${\bm h}^{(1)}(\bk)$ touch in a cone-like fashion. Moreover, also $h_z^{(1)}(\bk)=0$ for all $\bk$, 
since nearest-neighbor tunneling contributes only to the sublattice-mixing off-diagonal matrix elements. 
Now, the second-order correction to the effective Hamiltonian $\Ho_F^{(2)}$ contains complex next-nearest-neighbor 
tunneling matrix elements $-|J^{(2)}|e^{\pm i\vartheta}$ [see Fig.~\ref{fig:hexagonal}(b)] of the order of
$J^2/(\hbar\omega)$ [keeping $K/(\hbar\omega)$ fixed]. This gives rise to a correction ${\bm h}_F^{(2)}(\bk)$, where
only the $z$-component ${h}_{Fz}^{(2)}(\bk)$ is non-zero, since next-nearest-neighbor tunneling is sublattice preserving. 
Moreover, the time-reversal symmetry breaking associated with the phase $\vartheta$ makes this $z$-component non-zero at 
the Dirac points, such that ${h}_{Fz}^{(2)}(\bk_+)=-{h}^{(2)}_{Fz}(\bk_-)\ne0$. In this way the second-order correction 
removes the band touching and opens a gap between both bands. The opposite sign of ${h}^{(2)}_{Fz}(\bk_+)$  and
${h}^{(2)}_{Fz}(\bk_-)$ implies a finite Chern number of $\pm1$ of the lowest band \cite{Haldane88}. Now it is important 
to note that in second order the bands are separated by a gap of order
$J^2/(\hbar\omega)$ (excluding situations, where the forcing strength is fine-tuned close to values giving $J^{(1)}=0$ or
$J^{(2)}=0$). Even higher-order corrections, which in real space describe tunneling at even longer distances associated 
with a particle tunneling three or more times during one driving period, will be of the order of $J^3/(\hbar\omega)^2$.
They will not be able to close this gap and change the topological properties of the bands anymore. 
Thus, the essential physics of the driven system is captured by the approximation~(\ref{eq:Heff[2]}).

Qualitatively new behavior beyond the high-frequency approximation occurs, however, when $\hbar\omega$ becomes small 
enough so that $\varepsilon^F_+(\bk)-\varepsilon^F_-(\bk)=m\hbar\omega$ with $m=1,2,3,\dots$ for some quasimomenta
$\bk$. In this case both bands are coupled resonantly in an $m$-``photon'' process and hybridize. Such a hybridization is 
not captured by the perturbative approach underlying the high-frequency approximation. It occurs, roughly, when the 
gap between subspaces of different ``photon'' number closes. The new band gaps resulting from the avoided quasienergy 
level crossing between the states of quasienergy $\varepsilon^F_-(\bk)$ and $\varepsilon^F_+(\bk)-m\hbar\omega$ have been 
shown to give rise to intriguing topological properties without analog in non-driven systems \cite{KitagawaEtAl10,
RudnerEtAl13}.

\subsection{\label{sec:HexFM}Comparison with Floquet-Magnus expansion}

The circularly driven hexagonal lattice is also an instructive example that illustrates the difference 
between the high-frequency expansion of the effective Hamiltonian $\Ho_F$ on the one hand and of the Floquet 
Hamiltonian $\Ho_{t_0}^F$, as it appears in the Floquet-Magnus expansion, on the other. 

According to Eqs.~(\ref{eq:HFM1}) and (\ref{eq:HFM2}), the leading terms of the expansion of $\Ho^F_{t_0}$ read
\be
\Ho^{F(1)}_{t_0} =\Ho^{(1)}_F
\ee
and
\be\label{eq:HFM2hex}
\Ho^{F(2)}_{t_0} = \Ho^{(2)}_F + \sum_{m\ne0}\frac{1}{m\hbar\omega}e^{im\omega t_0} [\Ho_0,\Ho_m].
\ee
Evaluating the difference between the Floquet Hamiltonian and the effective Hamiltonian in 
second order, one obtains
\be\label{eq:HFM2hex}
\Ho^{F(2)}_{t_0} -\Ho^{(2)}_F
        =-\sum_{\lla\ell'\ell\rra} J^{(2)}_{t_0\lla\ell'\ell\rra} \,\aa_{\ell'}\ao_\ell.
\ee
Here
\bes\label{eq:J2t0} 
J^{(2)}_{t_0\lla\ell'\ell\rra}
    &=& - \sum_{m\ne0}\frac{J^2}{m\hbar\omega}\bJ_0\big({\T\frac{K}{\hbar\omega}}\big)
           \bJ_m\big({\T\frac{K}{\hbar\omega}}\big)  
     \bigg(e^{im(\omega t_0-\varphi_{k\ell})}-e^{im(\omega t_0-\varphi_{\ell'k})}\bigg) 
\nonumber\\&\simeq& 
    - \frac{2J^2}{\hbar\omega}\bJ_0\big({\T\frac{K}{\hbar\omega}}\big)\bJ_1\big({\T\frac{K}{\hbar\omega}}\big)
    \big[\cos(\omega t_0-\varphi_{k\ell})-\cos(\omega t_0-\varphi_{\ell'k})\big]
\nonumber\\
\ees
denotes the $t_0$-dependent part of the next-nearest-neighbor tunneling matrix element 
\be\label{eq:J2FMt0}
J^{(2)}_{\lla\ell'\ell\rra}+ J^{(2)}_{t_0\lla\ell'\ell\rra}
\ee
of $\Ho^{F(2)}_{t_0}$, with $k$ defined as the intermediate site between $\ell$ and $\ell'$.
It is easy to check that the tunneling matrix element (\ref{eq:J2FMt0}) depends on the direction of tunneling and not 
only on whether a particle tunnels in clockwise or anticlockwise direction around a hexagonal plaquette. This directional 
dependence is determined by $t_0$ and does not only concern the phase, but also the amplitude of next-nearest-neighbor 
tunneling matrix element (\ref{eq:J2FMt0}).
The consequence is a spurious $t_0$-dependent breaking of the discrete rotational symmetry of the band structure of the 
approximate Floquet Hamiltonian $\Ho_{t_0}^{F(1)}+\Ho_{t_0}^{F(2)}$. This can be seen as follows. 

\begin{figure}[t]
\includegraphics[width=1\linewidth]{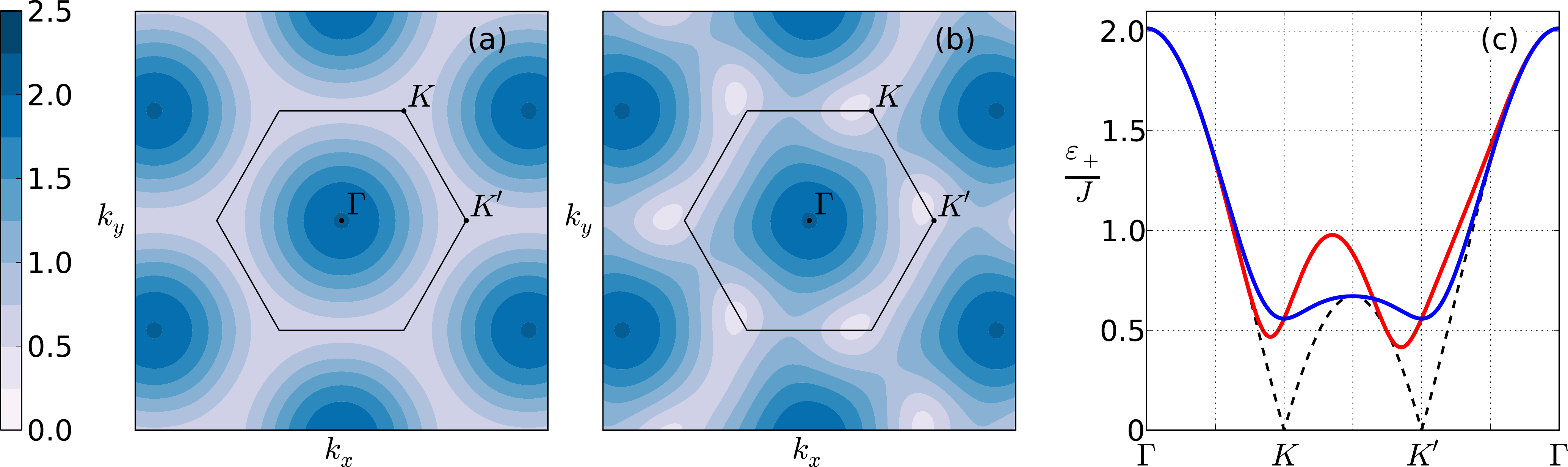}
\centering
\caption{\label{fig:dispersion} Quasienergy dispersion relation $\varepsilon_+(\bk)$ of the circularly driven hexagonal 
lattice, for $\hbar\omega/J=4$ and $K/(\hbar\omega)=1.2$. We compare the second-order high-frequency approximation of the 
effective Hamiltonian, $\varepsilon^{[2]}_+(\bk)$, [(a) and blue line in (c)] with the second-order Floquet-Magnus 
approximation, $\varepsilon^{FM[2]}_{t_0+}(\bk)$, evaluated for $\omega t_0=2\pi/3$ [(b) and red line in (c)]. The 
Floquet-Magnus approximation $\varepsilon^{FM[2]}_{t_0+}(\bk)$ does not only acquire a spurious $t_0$ dependence, but 
also breaks the discrete rotational symmetry of the exact dispersion relation $\varepsilon_+(\bk)$. In first order both 
approximations coincide, $\varepsilon^{FM[1]}_{t_0+}(\bk)=\varepsilon^{[1]}_{+}(\bk)$ [dashed line in~(c)].} 
\end{figure}

The driven Hamiltonian $\Ho(t)$ obeys the discrete translational symmetry of the hexagonal lattice so that quasimomentum 
is a conserved quantity. Therefore, $\Ho^{F}_{t_0}$ and $\Ho_F$ do not only possess the same spectrum, but 
also the same single-particle dispersion relation $\varepsilon_\pm(\bk)$. The symmetry of the hexagonal lattice with 
respect to discrete spatial rotations by $2\pi/3$ is broken by the periodic force. However, the force leaves the 
Hamiltonian $\Ho(t)$ unaltered with respect to the joint operation of a rotation by $2\pi/3$ combined with a time shift 
by $-T/3$. This spatio-temporal symmetry ensures that the effective Hamiltonian $\Ho_F$ possesses again the full discrete 
rotational symmetry of the hexagonal lattice. This is reflected in the leading terms (\ref{eq:H1hex}) and (\ref{eq:H2hex})
of the high-frequency expansion. As a consequence, the quasienergy band structure, given by the single particle 
dispersion relation $\varepsilon_\pm(\bk)$, is also symmetric with respect to a spatial rotation by $2\pi/3$.
The Floquet Hamiltonian $\Ho_{t_0}^{F}$, whose parametric dependence on the time $t_0$ indicates 
that it depends also on the micromotion, does not obey the discrete rotational symmetry of the hexagonal
lattice when evaluated for a fixed time $t_0$. However, the single-particle spectrum of $\Ho_{t_0}^{F}$ is still given by 
$\varepsilon_\pm(\bk)$ and, thus, rotationally symmetric. This latter property of the exact Floquet Hamiltonian
$\Ho^F_{t_0}$ is not preserved by the Floquet-Magnus approximation
$\Ho^{FM[2]}_{t_0}=\Ho_{t_0}^{F(1)}+\Ho_{t_0}^{F(2)}$. Here the second-order term does not only lead to a spurious
$t_0$ dependence of the spectrum, but also breaks the rotational symmetry of the quasienergy band structure. This is 
illustrated in Fig.~\ref{fig:dispersion}, where we compare the approximate dispersion relation $\varepsilon^{[2]}_+(\bk)$ 
resulting from the high-frequency approximation of the effective Hamiltonian with the spectrum
$\varepsilon^{FM[2]}_{t_0+}(\bk)$ obtained using the Floquet-Magnus approximation. In this respect, the approximate Floquet Hamiltonian
$\Ho^{FM[2]}_{t_0}$ is inconsistent with Floquet theory. The origin of this inconsistency is that the dispersion relation 
in Floquet-Magnus approximation depends on the driving phase, which, in turn, depends on the direction. Even though the 
spurious symmetry breaking should be small and of the order of $J (J/\hbar\omega)^{2}$ [for $(K/\hbar\omega)$ 
fixed], corresponding to the neglected third order, it still changes the property of the system in a fundamental way. 
Therefore, the Floquet-Magnus expansion has to be employed with care.

\subsection{Micromotion}

The micromotion operator $\Uo_F(t)=\exp\big(\hat{G}(t)\big)$ resulting from the periodic Hamiltonian $\Ho(t)$
is approximated by $\hat{G}(t)\approx \hat{G}^{(0)}(t)+\hat{G}^{(1)}(t)$. Here $\hat{G}^{(0)}(t)=0$ and, employing
Eq.~(\ref{eq:G(1)}), we find
\be
\hat{G}^{(1)}(t)= -\sum_{m=1}^\infty \frac{e^{im\omega t}\Ho_m-e^{-im\omega t}\Ho_{-m}}{m\hbar\omega}
=   \sum_{\la\ell'\ell\ra} g_{\ell'\ell}(t) \aa_{\ell'}\ao_\ell \;,
\ee
with
\bes\label{eq:gell}
g_{\ell'\ell}(t) &=& -\frac{J}{\hbar\omega} \sum_{m=1}^\infty \frac{1}{m}\bJ_m\big({\T\frac{K}{\hbar\omega}}\big)
    \Big[e^{im(\omega t-\varphi_{\ell'\ell})} - (-)^m e^{-im(\omega t-\varphi_{\ell'\ell})}\Big]
\nonumber\\&\simeq&
-\frac{2J}{\hbar\omega}\bJ_1\big({\T\frac{K}{\hbar\omega}}\big)\cos(\omega t-\varphi_{\ell'\ell}).    
\ees
Since $\varphi_{\ell\ell'}=\varphi_{\ell'\ell}+\pi$, it follows that
$g_{\ell'\ell}(t)=-g^*_{\ell\ell'}(t)$, such that $\hat{G}^{(1)}(t)$ is anti-hermitian as required.
With respect to the original frame of reference, where the system is described by the driven Hamiltonian
$\Ho_\text{dr}(t)$, the micromotion operator is given by 
\bes
\Uo_\text{dr}^F(t) &=& \Uo_F(t)\Uo(t)
\approx 
\exp\Bigg(\sum_{\la\ell'\ell\ra} g_{\ell'\ell}(t) \aa_{\ell'}\ao_\ell\Bigg)
        \exp\Bigg(i\sum_\ell\chi_\ell(t)\no_\ell\Bigg).
\nonumber\\
\ees

The dynamics that is described by $\Uo(t)$ does not happen in real space, but corresponds to a global 
time-periodic oscillation in quasimomentum by 
$\frac{K}{\hbar\omega}\frac{\hbar}{a}[\sin(\omega t){\bm e}_x-\cos(\omega t){\bm e}_y]$. 
This momentum oscillation is significant when $K\sim\hbar\omega$ and it is taken into account via the 
initial gauge transformation in a non-perturbative fashion, as can be seen from the Bessel-function-type 
dependence of the effective tunneling matrix elements on $K/\hbar\omega$. 
In turn, $\Uo_F(t)$ conserves quasimomentum and describes a micromotion in \emph{real space}. This real-space micromotion 
becomes significant, when the tunneling time $2\pi\hbar/J$ is not too large compared to the driving period $T$,
i.e.\ for $J/\hbar\omega$ not too small. A significant second-order correction $\Ho_F^{(2)}$ is a direct 
consequence of this real-space micromotion. In the high-frequency expansion, the real-space micromotion is
taken into account perturbatively. 

By expanding the tunneling matrix elements also in powers of the driving strength $K$, one finds that 
\be
\frac{J^{(2)}}{J^{(1)}} \simeq -\sqrt{3} \frac{ JK^2}{(\hbar\omega)^3}.
\ee
The quadratic dependence on $K$ indicates that large driving amplitudes $K/\hbar\omega\sim1$ are important in 
order to achieve a topological band gap $\sim|J^{(2)}|$ that is significant with respect to the band width
$\sim|J^{(1)}|$. At the same time, the linear dependence on the tunneling matrix element $J$ reveals that
moderate values of $J/\hbar\omega$, for which the high-frequency expansion is still justified, can be 
sufficient (see also Fig.~\ref{fig:Jeff}).

The time-dependent basis states capturing the micromotion can be constructed from the basis of Fock states
$|\{n_\ell\}\ra$ characterized by sharp on-site occupation numbers $n_\ell$. They read
\bes
|\{n_\ell\}(t)\ra_F &=& \Uo_F(t)|\{n_\ell\}\ra
\approx\exp\Bigg(\sum_{\la\ell'\ell\ra} g_{\ell'\ell}(t) \aa_{\ell'}\ao_\ell\Bigg)|\{n_\ell\}\ra
\nonumber\\
&\approx&\bigg(1
    -\sum_{\la\ell'\ell\ra}\frac{2J}{\hbar\omega}\bJ_1\big({\T\frac{K}{\hbar\omega}}\big)
        \cos(\omega t-\varphi_{\ell'\ell})  \aa_{\ell'}\ao_\ell\bigg)|\{n_\ell\}\ra,            
\ees
For the second approximation we have expanded the exponential, such that the state is normalized only 
up to terms of order $|J/\hbar\omega|^2$, and replaced $g_{\ell'\ell}(t)$ by the leading term of the sum
(\ref{eq:gell}). The micromotion is dominated by the oscillatory dynamics of particles from one lattice site 
to a neighboring one and back. The probability for a particle participating in such an oscillation is
of the order of $\big|\frac{2zJ}{\hbar\omega}\bJ_1\Big(\frac{K}{\hbar\omega}\Big)\big|^2$, with the 
coordination number $z=3$ counting the nearest neighbors of a lattice site. 

\section{\label{sec:interactions}Role of interactions}

If we consider a periodically driven quantum system of many particles, then the presence of interactions will 
influence the high-frequency expansion in two different ways. First, the interaction terms in the Hamiltonian 
will simply generate new terms in the high-frequency expansion. This effect will be discussed in the following 
subsection \ref{sec:IntCorr} for the example of the driven tight-binding lattice discussed in the 
previous section. Second, the presence of interactions will severely challenge the validity of the
high-frequency expansion, since even if the single-particle spectrum is bounded with a width lower than
$\hbar\omega$, this will not be the case anymore for collective excitations. This issue will be addressed in 
subsection \ref{sec:IntVal}.

\subsection{\label{sec:IntCorr}Interaction corrections within the high-frequency expansion}
Ignoring for the moment concerns that the high-frequency expansion should be employed with care in the 
presence of interactions, let us have a look at the new terms that will be generated by finite interaction 
terms in the Hamiltonian. We will focus on the example of the driven tight-binding lattice that was discussed 
on the single-particle level in the previous section. 

Starting from the interacting problem $\Ho_\text{dr}(t)+\Ho_\text{int}$ with a time-independent interaction 
term $\Ho_\text{int}$ of the typical density-density type, the interactions are not altered by the gauge 
transformation (\ref{eq:gauge}). Thus, after the gauge transformation the interacting lattice system is 
described by the driven Hubbard Hamiltonian
\be\label{HDBH}
\Ho(t) + \Ho_\text{int},
\ee
with the time-periodic single-particle Hamiltonian $\Ho(t)$ given by Eq.~(\ref{eq:gauge}).
The presence of the time-independent term $\Ho_\text{int}$, with trivial Fourier components 
\be
\Ho_{\text{int}\, m} = \delta_{m,0}\Ho_\text{int},
\ee 
will lead to additional terms in the high-frequency expansion of the effective Hamiltonian that we will 
denote by $\Ho^{(m)}_{F\text{int}}$. The leading contribution appears in the first order [Eq.~(\ref{eq:HF1})] and 
is given by the time-independent operator $\Ho_\text{int}$ itself,
\be
\Ho^{(1)}_{F\text{int}}=\Ho_\text{int}.
\ee
The second-order correction (\ref{eq:HF2}) vanishes, 
\be
\Ho^{(2)}_{F\text{int}}=0,
\ee
because all Fourier components $\Ho_{\text{int}\,m}$ with $|m|>0$ vanish. Therefore, the leading
correction involving the interactions appears in third order [Eq.~(\ref{eq:HF3})] and reads
\be\label{eq:Hint3}
\Ho^{(3)}_{F\text{int}} = \sum_{m=1}^\infty 
    \Bigg(\frac{\big[\Ho_{-m},[\Ho_\text{int},\Ho_m]\big]}{2(m\hbar\omega)^2}+\text{ h.c.} \Bigg).
\ee
Here $\Ho_m$ denotes a Fourier component of the kinetic part $\Ho(t)$ of the Hamiltonian given by
Eq.~(\ref{eq:Hm}) of the previous section and ``h.c.'' stands for ``hermitian conjugate''. The presence of 
interaction corrections beyond the leading order $\Ho^{(1)}_{F\text{int}}=\Ho_\text{int}$, as they result from real-space 
micromotion, was overlooked in a recent work investigating the possibility of stabilizing a
fractional-Chern-insulator-type many-body Floquet state with interacting fermions in the circularly driven hexagonal 
lattice \cite{GrushinEtAl14}. A recent study, which is based on the results presented here, investigates the 
impact of the correction (\ref{eq:Hint3}) on the stability of both fermionic and bosonic Floquet fractional Chern 
insulators \cite{AnisimovasEtAl15}. It is found that the correction tends to destabilize such a topologically ordered 
phase.

Note that the high-frequency expansion of the Floquet Hamiltonian $\Ho^F_{t_0}$, as it appears in the
Floquet-Magnus expansion, produces also a second-order correction (\ref{eq:HFM2}) that involves the 
interactions. It is given by \cite{VerdenyEtAl13,BukovEtAl14,DAlessio14}
\be
\Ho^{F(2)}_{\text{int}, t_0}=\sum_{m=1}^\infty \frac{1}{m\hbar\omega}e^{im\omega t_0}[\Ho_\text{int},\Ho_m].
\ee
However, as we have discussed in section~\ref{sec:FM}, if we approximate the Floquet Hamiltonian in second 
order this term will not influence the many-body spectrum within the order of the approximation, while it 
introduces an unphysical dependence of the spectrum on the driving phase.

In order to get an idea of what type of terms appear in the high-frequency expansion, let us write down the 
third-order interaction correction for the simple case of spinless bosons in 
the circularly forced hexagonal lattice with on-site interactions
\be
\Ho_\text{int}=\frac{U}{2}\sum_\ell\no_\ell(\no_\ell-1).
\ee
This model system provides a quantitative description of ultracold bosonic atoms in an optical lattice and is 
interesting also because it might (as well as the fermionic system \cite{GrushinEtAl14}) be a possible 
candidate for a system stabilizing a Floquet fractional Chern insulator state. 
Using Eq.~(\ref{eq:Hint3}), we find those third-order correction terms that involve the interactions to be 
given by
\bes
\Ho^{(3)}_{F\text{int}} &=& 
           - \sum_\ell 2zW_a^{(3)}\no_\ell(\no_\ell-1)
\nonumber\\&&           
+\, \sum_{\la\ell'\ell\ra} 
            \Big\{4W_a^{(3)}\no_{\ell'}\no_\ell+ 2 W_b^{(3)}\aa_{\ell'}\aa_{\ell'} \ao_\ell\ao_\ell \Big\}
\nonumber\\&& 
         +\,\sum_{\la\ell'k\ell\ra} 
        \Big\{W^{(3)}_{c} \aa_{\ell'}\big(4\no_{k}-\no_{\ell}-\no_{\ell'} \big)\ao_{\ell}
\nonumber\\&&\qquad
       +\,W^{(3)}_{d} \big(\aa_k\aa_k\ao_{\ell'}\ao_\ell + \aa_\ell\aa_{\ell'}\ao_k\ao_k\big)\Big\}\bigg],
\ees
with coordination number $z=3$ and coupling strengths
\bes
  W_a^{(3)} &=&  \frac{UJ^2}{(\hbar\omega)^2} \sum_{m=1}^\infty 
    \frac{1}{m^2}\bJ_m^2\big({\T\frac{K}{\hbar\omega}}\big) 
    \simeq  \frac{UJ^2}{(\hbar\omega)^2}\bJ_1^2\big({\T\frac{K}{\hbar\omega}}\big),
\\
  W_b^{(3)}  &=& \frac{UJ^2}{(\hbar\omega)^2} \sum_{m=1}^\infty 
    \frac{(-)^{m+1}}{m^2} \bJ_m^2\big({\T\frac{K}{\hbar\omega}}\big) 
    \simeq  \frac{UJ^2}{(\hbar\omega)^2}\bJ_1^2\big({\T\frac{K}{\hbar\omega}}\big),
\\
  W^{(3)}_{c}
    &=& \frac{UJ^2}{(\hbar\omega)^2} \sum_{m=1}^\infty  
    \frac{1}{m^2} \cos \left(\frac{2\pi m}{3}\right) 
    \bJ_m^2\big({\T\frac{K}{\hbar\omega}}\big) 
    \simeq - \frac{UJ^2}{2(\hbar\omega)^2}\bJ_1^2\big({\T\frac{K}{\hbar\omega}}\big),
\\
W^{(3)}_{d}
    &=& \frac{UJ^2}{(\hbar\omega)^2} \sum_{m=1}^\infty 
    \frac{(-)^{m+1}}{m^2} \cos \left(\frac{2\pi m}{3}\right) 
    \bJ_m^2\big({\T\frac{K}{\hbar\omega}}\big) 
    \simeq - \frac{UJ^2}{2(\hbar\omega)^2}\bJ_1^2\big({\T\frac{K}{\hbar\omega}}\big).
\nonumber\\
\ees
The third sum runs over all directed three-site strings $\la\ell'k\ell\ra$, defined such that $\ell$ and $k$ 
as well as $k$ and $\ell'$ are nearest neighbors, while $\ell\ne\ell'$. 
The first sum produces a correction that reduces the on-site interactions of the leading term
$\Ho^{(1)}_{F\text{int}}=\Ho_\text{int}$. The second sum introduces both nearest-neighbor density-density 
interactions like in the extended Hubbard model and a pair-tunneling term. And the third sum, finally 
describes density-assisted tunneling between next-nearest neighbors, as well as the joint tunneling of
two-particles into or away from a given site.

\subsection{\label{sec:IntVal}On the validity of the high-frequency approximation for many-body systems}

The high-frequency expansion of the effective Hamiltonian $\Ho_F$ and the micromotion operator $\Uo_F(t)$ can 
generally not be expected to converge for a system of many interacting particles. Namely, the time average
$\Ho_0$ of the full many-body Hamiltonian $\Ho(t)$, which determines the spectrum of the diagonal blocks of 
the quasienergy operator $\bar{Q}$ depicted in Fig.~\ref{fig:quasienergy}, will possess collective 
excitations also at very large energies. Therefore the energy gaps of $\hbar\omega$, which separate the 
subspaces of different ``photon'' numbers $m$ in the unperturbed problem $\bar{Q}_0$, will close when the 
perturbation is switched on (unless $\hbar\omega$ was a macroscopic energy).

\begin{figure}[t]
\includegraphics[width=1\linewidth]{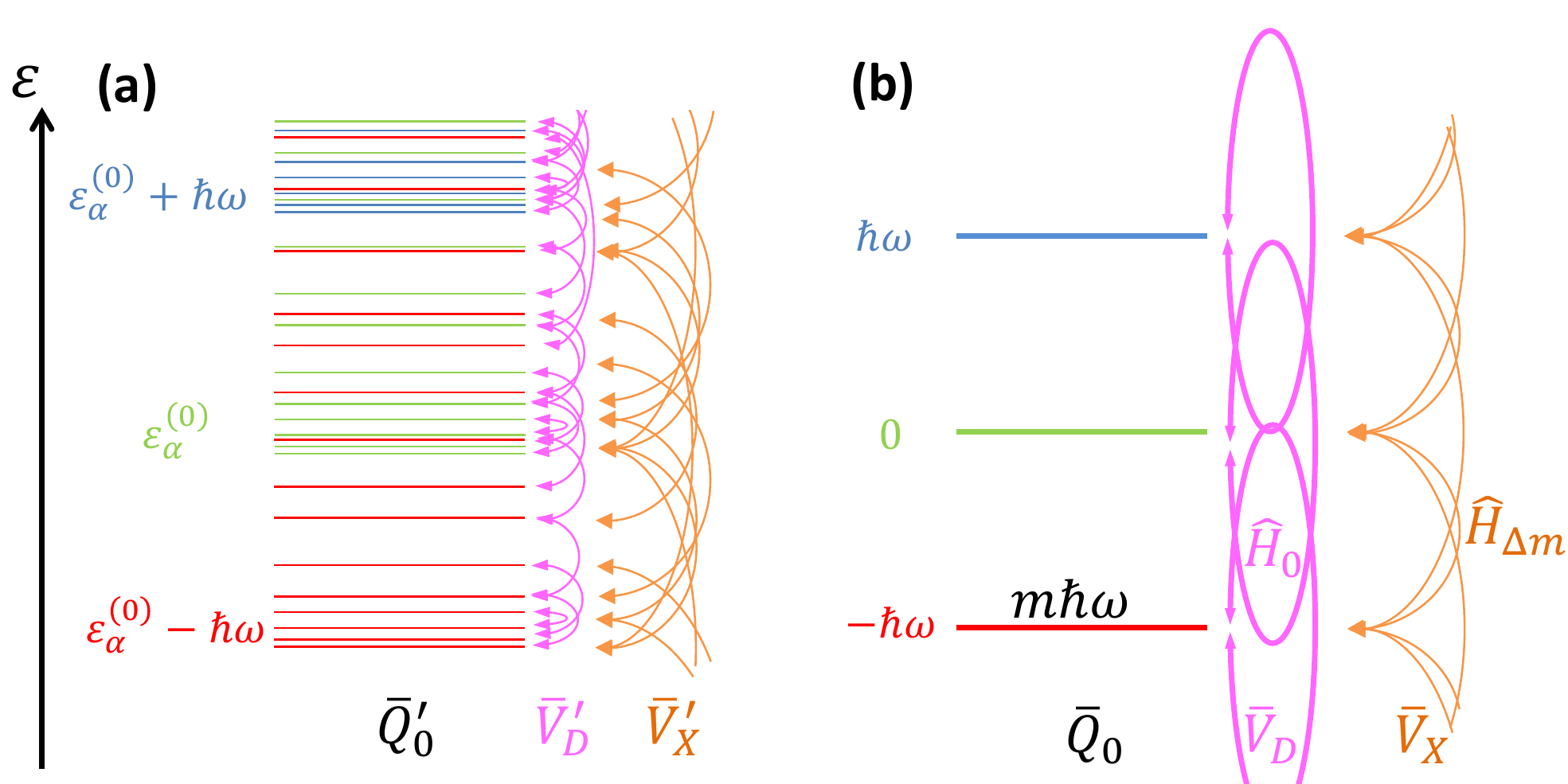}
\centering
\caption{\label{fig:NonQDeg}Like Fig.~\ref{fig:QDeg}, but with the spectral width defined by
$\varepsilon^{(0)}_\alpha$ larger than the energetic separation $\hbar\omega$ of sectors with different 
``photon'' number $m$. The perturbation expansion is not justified and cannot be expected to converge. However, defining 
the unperturbed problem like in subfigure (b), the perturbation expansion can still be written down formally and, under 
certain conditions, still provide a useful approximation.}
\end{figure}

Nevertheless, the fact that the unperturbed problem $\bar{Q}_0$ is given by the ``photonic'' part of the 
quasienergy operator, with exactly degenerate eigenvalues $m\hbar\omega$ in each subspace of photon number $m$,
allows one to formally write down the perturbation expansion. The energy denominators will not diverge. This 
is illustrated in Fig.~\ref{fig:NonQDeg}. And, even if the perturbation expansion can generally not be expected to 
converge, these terms can still provide an approximate description of the driven many-body system on a finite time scale. 
This can be the case if the $\nu$th-order approximate effective Hamiltonian $\Ho_F^{[\nu]}$ is 
governed by energy scales that are small compared to $\hbar\omega$. Then the creation of a collective 
excitation of energy $\hbar\omega$ corresponds to a significant change in the structure of the many-body wave 
function and is a process associated with a very small matrix element only.
As a consequence, on time scales that are small compared to the inverse of such residual matrix elements, the approximate 
effective Hamiltonian can be employed to compute the dynamics and approximate Floquet states of the system. This is
the basis for Floquet engineering in interacting systems.

Let us illustrate this rather abstract reasoning by a concrete example. Consider again spinless bosons in the circularly 
forced hexagonal lattice, described by the driven Bose-Hubbard Hamiltonian (\ref{HDBH}). The first-order approximation to 
the effective Hamiltonian is given by
\be\label{eq:HF1int}
\Ho_F^{(1)}= - J\bJ_0\big({\T\frac{K}{\hbar\omega}}\big)\sum_{\la\ell'\ell\ra}\aa_{\ell'}\ao_\ell
    +\frac{U}{2}\sum_\ell \no_\ell(\no_\ell-1).
\ee
In the thermodynamic limit (where the number of lattice sites $M$ is taken to infinity, while keeping the filling of $n$ 
particles per lattice site fixed) the system possesses excited states at arbitrarily large energies. Therefore, the width 
of the spectrum is obviously larger than $\hbar\omega$. As a consequence, the first-order quasienergy spectrum that
describes the subspace of ``photon'' number $m$ is much wider than $\hbar\omega$. It is not separated by a spectral gap
from subspaces of different ``photon number'' anymore, but overlaps with quasienergies from these spectra. Therefore, the 
perturbation expansion, which is based on the assumption that the unperturbed spectral gaps do not close in response to 
the perturbation, can generally not be expected to converge.\footnote{Note that there are exceptions to this expectation. 
For example, the non-interacting system with $U=0$ is described by a quadratic Hamiltonian. In this case the problem can be
reduced to a single-particle problem. As a consequence, convergence is expected already for $\hbar\omega$ sufficiently 
large compared to the width of the \emph{single-particle} spectrum $\sim J$. This is true even if
$\hbar\omega$ is small compared to the width of the $N$-particle spectrum $\sim NJ$ of the Hamiltonian (\ref{eq:HF1int}) 
with $U=0$. Namely, the matrix elements for the joint creation of several single-particle excitations at once vanish. 
However, if the particles are coupled by a finite interaction strength $U$, the single-particle picture does not apply 
anymore.} Namely, if we would consider the full diagonal blocks of the quasienergy matrix
(Fig.~\ref{fig:quasienergy}) as unperturbed problem, that is if we would add the block-diagonal part of the
supposedly perturbation to the unperturbed quasienergy operator considered before, already the unperturbed 
quasienergy spectra of different subspaces $m$ would overlap. As a consequence the energy denominators appearing in the 
perturbation expansion would diverge, whenever a degeneracy between unperturbed states of different ``photon'' number $m$
occurs. 

We can now identify processes that spoil the high-frequency expansion for the driven Bose-Hubbard model (\ref{HDBH}) and 
estimate the time scale on which they occur. For simplicity, we focus on the limit of strong interactions $U\gg J$ and 
assume a filling of $n=1$. In the spirit of the argument presented at the end of the preceding paragraph, we can now add 
part of the perturbation to the unperturbed problem. We include the interactions in the definition of the unperturbed 
quasienergy operator, 
\be
\Qo'_0= \Ho_\text{int}-i\hbar\rd_t,
\ee
so that the perturbation is given by the kinetic part (\ref{eq:gauge}) of the Hamiltonian,
\be
\Vo'(t)=\Ho(t)=\sum_m\Ho_m e^{im\omega t},
\ee
with Fourier components $\Ho_m$ given by Eq.~(\ref{eq:Hm}). 
The unperturbed Floquet states $|\{n_\ell\}m\rra$ are characterized by sharp site occupation numbers $n_\ell$ as well as 
sharp ``photon'' numbers $m$ and read
\be
|\{n_\ell\}m(t)\ra =e^{im\omega t}|\{n_\ell\}\ra.
\ee
They diagonalize the unperturbed problem, 
\be
\lla\{n'_\ell\}m'|\Qo'_0|\{n_\ell\}m\rra = \delta_{m',m}\delta_{\{n'_\ell\},\{n_\ell\}}\varepsilon'^{(0)}_{\{n_\ell\}m},
\ee
where the unperturbed quasienergies read
\be
\varepsilon'^{(0)}_{\{n_\ell\}m} = \frac{U}{2}\sum_\ell n_\ell(n_\ell-1) + m\hbar\omega.
\ee
The high-frequency approximation is spoiled whenever two unperturbed states $|\{n_\ell\}m\rra$ and $|\{n'_\ell\}m'\rra$ 
with different photon numbers $m\ne m'$ are nearly degenerate and coupled to each other (either directly or in a
higher-order process). In such a situation a strong hybridization between these states of different ``photon'' number 
occurs, rather than a perturbative admixture of one state to the other as required by the high-frequency approximation. 

Let us investigate in how far the unperturbed ``ground state'' of the $m=0$ manifold, $|\psi_0 \rra$, suffers from such 
detrimental coupling. This state corresponds to a Mott-insulator with one particle per site, 
\be
|\psi_0 \rra=|\{n_\ell=1\}0\rra.
\ee
It is the only unperturbed $m=0$ state without any doubly occupied site, so that the interaction energy vanishes. Its 
quasienergy reads $\varepsilon'^{(0)}_0=0$. It is the approximate ground state of the approximate effective Hamiltonians
(\ref{eq:HF1int}) in the regime $U\gg J$ \cite{FisherEtAl89}. We can now systematically study the most relevant 
processes that couple $|\psi_0\rra$ to states of ``photon'' number $m\ne0$. For that purpose, we will follow the 
procedure applied to a one-dimensional chain in reference \cite{EckardtHolthaus08b}. 

The state $|\psi_0\rra$ is coupled directly (in first-order\footnote{This order does not refer to the high-frequency 
approximation discussed so far, but to a perturbative approach treating $|\psi_0 \rra$ and collective excitations with
$m<0$ photons to be nearly degenerate.}) to unperturbed states 
\be
|\la\ell_2\ell_1\ra m\rra \equiv\frac{1}{\sqrt{2}}\ba_{\ell_2}\bo_{\ell_1} |\{n_\ell=1\}m\rra,
\ee
with one particle-hole excitation on neighboring sites $\ell_1$ and $\ell_2$, $m$ photons, and
unperturbed quasienergy $\varepsilon^{(0)}_{\la\ell_2\ell_1\ra m}=U+m\hbar\omega$. 
The resonance condition $\varepsilon^{(0)}_{\la\ell_2\ell_1\ra m}=\varepsilon'^{(0)}_0$ is fulfilled for frequencies
\be\label{eq:res1}
\hbar\omega=-\frac{U}{m}\qquad\text{with}\qquad
m=-1,-2,-3,\ldots.
\ee
They are labeled by the negative change of the ``photon'' number, $m$, associated with the transition
$|\psi_0\rra\to|\la\ell_2\ell_1\ra m\rra$. The corresponding coupling matrix elements is of the order of $J$ and reads
\bes
C^{(1)}_m=\lla \la\ell_2\ell_1\ra m|\bar{V}'|\psi_0\rra &=& 
    \frac{1}{\sqrt{2}}\la\{n_\ell=1\}|\ba_{\ell_1}\bo_{\ell_2}\Ho_m |\{n_\ell=1\}\ra 
\nonumber\\
    &=& -\sqrt{2}J\bJ_{m}\big({\T\frac{K}{\hbar\omega}}\big)e^{im\varphi_{\ell_2\ell_1}}.
\ees
The time scale for the resonant creation of a particle-hole excitations in the effective ground state can thus be 
estimated to be given by $2\pi\hbar/C^{(1)}_m$. If $\hbar\omega$ is larger than $U$, the direct excitation of
particle-hole pairs is not resonant, so that this type of heating process cannot spoil the high-frequency approximation. 

The next-strongest type of process is the creation of a collective excitation consiting of two coupled (i.e.\ overlapping)
particle-hole pairs, with two extra particles on a site $\ell_3$ and no particles on two neighboring sites $\ell_1$ and
$\ell_2$.\footnote{The coupling matrix element for the creation of two or more independent (i.e.\ non-overlapping) 
particle-hole pairs, corresponding to an unconnected diagram, can be shown to vanish. Otherwise the overall rate for the 
creation of isolated particle-hole pairs would grow in a superlinear (i.e.\ superextensive) fashion with the system size.}
Such a state is given by 
\be\label{eq:2ph}
|(\la\ell_3\ell_1\ra\la\ell'\ell_2\ra) m\rra 
=\frac{1}{\sqrt{6}}\ba_{\ell_3}\ba_{\ell_3}\bo_{\ell_1}\bo_{\ell_2} |\{n_\ell=1\}m\rra
\ee
and possesses the unperturbed quasienergy $\varepsilon^{(0)}_{(\la\ell_3\ell_1\ra\la\ell_3\ell_2\ra) m}=3U+m\hbar\omega$. 
The resonance condition $\varepsilon^{(0)}_{(\la\ell_3\ell_1\ra\la\ell_3\ell_2\ra) m}=\varepsilon'^{(0)}_0$ is fulfilled 
for frequencies
\be
\hbar\omega=-\frac{3U}{m}\qquad\text{with}\qquad
m=-1,-2,-4,-5,-7\ldots. 
\ee
Here we have omitted those $m$, for which already a dominating first-order resonance (\ref{eq:res1}) occurs.
The state (\ref{eq:2ph}) is coupled to $|\psi_0\rra$ only indirectly, in a second-order process via intermediate states 
$|\la\ell_3\ell_{1,2}\ra m'\rra$ characterized a single particle-hole pair and photon number $m'$. The coupling matrix 
element can be evalutated using degenerate perturbation theory and reads
\bes
C^{(2)}_m &=& \sum_{m'=-\infty}^\infty  
		\lla (\la\ell_3\ell_1\ra\la\ell_3\ell_2\ra) m|\bar{V}
		\bigg(\sum_{\ell=\ell_1,\ell_2}  |\la\ell_3\ell\ra m'\rra \lla \la\ell_3\ell\ra m'|\bigg)\bar{V}|\psi_0\rra
\nonumber\\&& \times\frac{1}{2}\Bigg[
	\frac{1}{\varepsilon^{(0)}_{(\la\ell_3\ell_1\ra\la\ell_3\ell_2\ra) m}-\varepsilon^{(0)}_{\la\ell_3\ell_{1,2}\ra m'}}
	+\frac{1}{\varepsilon^{(0)}_{0}-\varepsilon^{(0)}_{\la\ell_3\ell_{1,2}\ra m'}} \Bigg] 
\nonumber\\&=&
	\frac{\sqrt{6}J^2}{\hbar\omega}
	\sum_{m'}\frac{\bJ_{m-m'}\big({\T\frac{K}{\hbar\omega}}\big)\bJ_{m'}\big({\T\frac{K}{\hbar\omega}}\big)}{m/3-m'}
\nonumber\\&& \times
	\bigg(e^{i(m-m')\varphi_{\ell_3\ell_2} + im'\varphi_{\ell_3\ell_1}}
		+e^{i(m-m')\varphi_{\ell_3\ell_1} + im'\varphi_{\ell_3\ell_2}} \bigg).
\ees
In order to arrive at a simple result, we have focused on the vicinity of the resonance and employed $U=-m\hbar\omega/3$ 
in the energy denominators. The time scale for the resonant creation of such an overlapping pair of particle-hole 
excitations is given by $2\pi\hbar/C^{(2)}_m$. It is increased by a factor of the order of $\hbar\omega/J$ with respect to the creation of a 
single particle-hole pair, since it appears one order higher in perturbation theory. The experimentalists can avoid 
heating processes associated with the creation of two coupled particle-hole excitations by choosing $\hbar\omega$ well 
above $3U$. In this case, the most detrimental process will be the creation of a collective excitation given by three 
overlapping particle-hole pairs in a third-order process, with matrix element $C^{(3)}_m\sim J^3/(\hbar\omega)^2$. These 
third-order heating processes can, in turn, be avoided for $\hbar\omega$ well above $12U$, and so on. Thus, by 
increasing the driving frequency, the time scale for detrimental heating due to the creation of collective excitations 
increases.

The order of perturbation theory in which detrimental heating processes beyond the high-frequency approximation occur 
increases like a power law with the frequency $\hbar\omega$, at least for the model and the parameters considered. This 
suggests that the time scale for heating due to the creation of collective excitations increases exponentially with the 
driving frequency (the higher the energy of a collective excitation the more complex it is and the smaller will be the 
matrix element for its creation). Such a scaling is favorable for quantum engineering based on the high-frequency 
approximation. However, one has to note that the driving frequency cannot simply be increased to arbitrarily large values,
because with increasing driving frequency another type of heating process will become more relevant. Namely, the resonant 
creation of high-energy single-particle states neglected in a low-energy model will become more and more relevant. In the 
driven lattice system these states belong to excited Bloch bands above a large energy gap $\Delta$ that 
are not taken into account in the tight binding description. If the resonance condition $\Delta \approx m\hbar \omega$ is 
fulfilled, these states can be populated in $m$-photon processes \cite{ArlinghausHolthaus12,BilitewskiCooper15,
WeinbergEtAl15}, the time scale of which tends to increase with $m$. Thus, Floquet engineering requires a window of 
suitable driving frequencies that are both large enough to suppress heating due to the creation of collective excitations 
and small enough to suppress heating due to interband transitions. In optical lattice systems, the existence of a window 
of suitable driving frequencies has been demonstrated in a number of experiments (see first paragraph of section
\ref{sec:introduction}).  

In the preceding paragraph the requirement ``to suppress heating'' should be read as ``to suppress heating on the 
time scale of the experimental protocol''. Thus, the window of suitable frequencies depends also on the duration of the 
experiments, which in turn is determined by the effects to be studied. This dependence on the duration of the experiment 
is reflected also, in the response of a system to parameter variations. An important experimental protocol in the context
of Floquet engineering is, for example, the smooth switching-on of the driving amplitude. Here the aim is to start from the
ground state of the undriven model and to adiabatically prepare the ground state of the effective Hamiltonian obtained 
using the high-frequency approximation. In order to understand such parameter variations, one can employ the adiabatic 
principle for Floquet states \cite{BreuerHolthaus89b}. In order to achieve the desired dynamics in a driven many-body 
system, it is not only required that the parameter variation is slow enough to ensure adiabatic following with 
respect to the high-frequency approximate effective Hamiltonian. At the same time, the parameter variation must also be 
sufficiently \emph{fast}. Namely the matrix elements for the resonant creation of collective excitations discussed above 
will lead to tiny avoided crossings in the quasienergy spectrum between states of different ``photon'' number. These 
anticrossings are not captured by the high-frequency approximation and, in order to avoid heating, have to be passed 
diabatically \cite{EckardtHolthaus08b}. 
 
The window of suitable driving frequencies is not necessarily determined by heating processes alone. Another limitation 
comes in, if the second-order term of the high-frequency approximation, $\Ho_F^{(2)}$, is relevant for the model system 
to be realized via Floquet engineering. The opening of a topological band gap in the circularly driven hexagonal lattice 
discussed in section \ref{sec:hexHeff} is a prime example. This band gap appears in second order and depends inversely on 
the driving frequency $\omega^{-1}$ [if $K/(\hbar\omega)$ is held constant]. Thus if the band gap is required to be 
larger than some minimum value (determined, e.g., by interactions, temperature, or the rate of a parameter variation), 
this sets an upper limit for the driving frequency. The existence (and the identification) of system parameters for which 
a window of suitable driving frequencies exists is an important issue of Floquet engineering. 

The reasoning of this subsection suggests that, when dealing with a large system of interacting particles, on long time 
scales the high-frequency approximation can be expected to break down due to heating. In fact, the expected generic 
behavior of a driven many-body system in the thermodynamic limit is that it eventually approaches an
infinite-temperature-like state in the long-time limit \cite{LazaridesEtAl14b,DAlessioRigol14}. Though, exceptions are 
conjectured to exist \cite{LazaridesEtAl14a,AbaninEtAl14,LazaridesEtAl14c,CitroEtAl15,PonteEtAl15,PonteEtAl15b,
RussomannoEtAl15}, including integrable systems and systems featuring many-body localization, where the size of the many-
body state space is effectively reduced via the segmentation into different subspaces. In the sense of eigenstate 
thermalization \cite{Srednicki94,RigolEtAl08}, the conjectured infinite-temperature behavior is attributed to the resonant
($m$-``photon''-type) coupling, and the resulting hybridization, of states with very different mean energies. However, 
irrespective of the long-time behavior, the high-frequency approximation might still provide an accurate description of a 
driven many-body system on a finite time scale.

\section{\label{sec:conclusions}Summary}
We have used degenerate perturbation theory in the exended Floquet space to derive a high-frequency expansion of 
the effective Hamiltonian and the micromotion operator of periodically driven quantum systems. This approach provides an 
intuitive picture of the nature of the high-frequency approximation and its limitations. We have, moreover, related our 
approach to the Floquet-Magnus expansion. We have discussed that the latter is plagued not only by a spurious $t_0$
dependence of the quasienergy spectrum, but that it can also violate further symmetries of the exact result, like the 
rotational symmetry of a Floquet band structure. Finally, we have addressed the possibly detrimental impact of 
interactions on the validity of the high-frequency approximation beyond a certain time scale. 

\section*{Acknowledgements}
We thank Adolfo Grushin, Gediminas Juzeli\={u}nas, Achilleas Lazarides, and Titus Neupert for discussion. 
This research was supported by the European Social Fund under the Global Grant measure.

\begin{appendix}
\section{Existence of Floquet states\label{sec:existence}}
One can show that the eigenstates of the time-evolution operator over one period have the properties of 
Floquet states. The eigenstates $|\psi_n(t)\ra$ of the time evolution operator from time $t$ to time
$t+T$ fulfill
\be\label{eq:eigen}
\Uo(t+T,t)|\psi_n(t)\ra= a_n(t)|\psi_n(t)\ra.
\ee
Since $U(t+T,t)$ is unitary, the eigenvalues $a_n(t)$ are phase factors, $|a_n(t)|=1$, and the eigenstates
$|\psi_n(t)\ra$ can be chosen to form a complete orthogonal basis. It is left to show that the
time-evolved states $|\psi_n(t')\ra=\Uo(t',t)|\psi(t)\ra$ are eigenstates of $\Uo(T+t',t')$ with the 
same eigenvalue, i.e.\ that $a_n(t')=a_n(t)\equiv a_n\equiv e^{-i\varepsilon_n T/\hbar}$. For that purpose one 
introduces $1=\Uo(t,t')\Uo(t',t)$ on both sides of the eigenvalue equation
(\ref{eq:eigen}) and multiplies the equation by $\Uo(t',t)$ from the left. Furthermore, the periodic
time-dependence of the Hamiltonian has to be employed by using $\Uo(t',t)=\Uo(t'+T,t+T)$. One arrives at 
\be
\Uo(t'+T,t')|\psi_n(t')\ra= a_n(t)|\psi_n(t')\ra,
\ee
such that, indeed, $a_n(t')=a_n(t)$. Thus, one has
\be
\Uo(t+T,t)|\psi_n(t)\ra= e^{-i\varepsilon_n T/\hbar}|\psi_n(t)\ra,
\ee
with real quasienergy $\varepsilon_n$ not depending on the time $t$. The Floquet states
$|\psi_n(t)\ra$ fulfill 
\be
|\psi_n(t+T)\ra=e^{-i\varepsilon_n T/\hbar}|\psi_n(t)\ra
\ee
and can be written like
\be
|\psi_n(t)\ra= e^{-i\varepsilon_n t/\hbar}|u_n(t)\ra ,
\ee
where the time-periodic Floquet mode, 
\be
|u_n(t)\ra  =  e^{i\varepsilon_n T/\hbar}|\psi_n(t)\ra =|u_n(t+T)\ra,
\ee
has been introduced.

\section{Relations between operators in $\mathcal{H}$ and $\mathcal{F}$ \label{sec:operators}}

An important class of operators $\bar{A}$ acting in $\mathcal{F}$ are those corresponding to operators
$\Ao(t)=\Ao(t+T)$ that act in $\mathcal{H}$ and possess a periodic and local time dependence. Here 
local in time means that the operator does neither involve a time derivative nor an integral over a finite time span. Let us briefly summarize some properties of such operators. 

(i) The operator $\bar{A}$ corresponding to $\Ao(t)$ possesses matrix elements
\be\label{eq:Ame}
\lla\alpha'm'|\bar{A}|\alpha m\rra = \la\alpha'|\Ao_{m'-m}|\alpha\ra
\ee
that are determined by the Fourier components 
\be
\Ao_m=\frac{1}{T}\int_0^T \! \rd t\, e^{-im\omega t}\Ao(t)
\ee
of $\Ao(t)$. The fact that the matrix elements (\ref{eq:Ame}) depend on the difference $m'-m$ show that 
time-periodic time-local (TPTL) operators $\Ao(t)$ correspond to operators $\bar{A}$ that are translation 
invariant with respect to the ``photon'' number $m$. 
Vice versa, for every operator $\bar{A}$ being translationally invariant with respect to $m$, we can 
construct a corresponding time periodic operator 
\be\label{eq:FtoH}
\Ao(t) = \sum_m \Ao_m e^{im\omega t}
\ee
acting in $\mathcal{H}$ via
\be\label{eq:AmF}
\Ao_m=\sum_{\alpha',\alpha} |\alpha'\ra \lla \alpha'm|\bar{A}|\alpha 0\rra\la\alpha|.
\ee
An example for such a TPTL operator is the Hamiltonian $\Ho(t)$, giving rise to matrix elements
$\lla\alpha'm'|\bar{H}|\alpha m\rra= \la\alpha'|\Ho_{m'-m}|\alpha\ra$. An example for an 
operator that is not time-local is the ``photonic'' part $\Qo_p(t)=-i\hbar\rd_t\hat{1}$ of the quasienergy 
operator. Here we have explicitly written out the 
unity operator $\hat{1}$ in $\mathcal{H}$, which we usually suppress. The corresponding operator
$\bar{Q}_p$ in $\mathcal{F}$ possesses matrix elements
$\lla\alpha'm'|\bar{Q}_p|\alpha m\rra=\delta_{m'm}\delta_{\alpha'\alpha} m\hbar\omega$ that are not 
translationally invariant with respect to the ``photon'' number.

(ii) A TPTL operator such as the Hamiltonian $\Ho(t)$ that for all times $t$ is 
hermitian in $\mathcal{H}$ corresponds to an operator $\bar{H}$ in $\mathcal{F}$ that is 
translationally invariant with respect to the ``photon'' number $m$ and hermitian, and vice versa,
\be
\Ho^\dag(t)=\Ho(t) \qquad \Leftrightarrow\qquad \bar{H}^\dag=\bar{H}.
\ee
For example, one direction (``$\Rightarrow$'') can be shown as follows:
$\lla\alpha'm'|\bar{H}^\dag|\alpha m\rra = \lla\alpha m|\bar{H}|\alpha' m'\rra^* = 
\la\alpha|\Ho_{m-m'}|\alpha'\ra^*=\la\alpha'|\Ho_{m-m'}^\dag|\alpha\ra 
=\la\alpha'|\Ho_{m'-m}|\alpha\ra = \lla\alpha'm'|\bar{H}|\alpha m\rra$. Here we have employed
$\Ho_{m-m'}^\dag=-\Ho_{m'-m}$, which follows from $\Ho(t)=\sum_me^{im\omega t}\Ho_m=\Ho^\dag(t)$.

(iii) A multiplication of two TPTL operators in $\mathcal{H}$ directly 
corresponds to a multiplication in $\mathcal{F}$, 
\be
\Ao(t)=\Bo(t)\Co(t) \qquad\Leftrightarrow\qquad \bar{A}=\bar{B}\bar{C}.
\ee
The proof is straightforward. This implies also that
\be\label{eq:function}
\Ao(t)=f\big(\Bo(t)\big) \qquad\Leftrightarrow\qquad \bar{A}=f\big(\bar{B}\big),
\ee
where the function $f$ is defined via its Taylor expansion.  

(iv) When viewed as a constant function in time, the unity operator $\hat{1}$ in $\mathcal{H}$, with
$\la\alpha'|\hat{1}|\alpha\ra=\delta_{\alpha',\alpha}$, directly corresponds to the unity operator
$\bar{1}$ in $\mathcal{F}$, with $\lla\alpha'm'|\bar{1}|\alpha m\rra=\delta_{m'm}\delta_{\alpha'\alpha}$.
That is 
\be
\Ao(t)=\hat{1} \qquad\Leftrightarrow\qquad \bar{A} =\bar{1}.
\ee

(v) A TPTL operator $\Uo(t)$ that for all times $t$ is unitary in $\mathcal{H}$ corresponds to an operator $\bar{U}$
in $\mathcal{F}$ that is translationally invariant with respect to the ``photon'' number $m$ and unitary, and vice versa,
\be
\Uo^\dag(t)\Uo(t)=\Uo(t)\Uo^\dag(t) =\hat{1}
    \qquad \Leftrightarrow\qquad 
\bar{U}^\dag\bar{U}=\bar{U}\bar{U}^\dag=\bar{1}.
\ee
This is a direct consequence of (iii) and (iv).

\section{\label{sec:perturbation} Degenerate perturbation theory in the extended Floquet Hilbert space}

Degenerate perturbation theory is an approximation scheme that allows for the systematic block 
diagonalization of a hermitian operator into two subspaces separated by a spectral gap. Here 
we apply the canonical van Vleck degenerate perturbation theory \cite{ShavittRedmon80} to the 
quasienergy operator $\bar{Q}$ in the extended Floquet Hilbert space $\mathcal{F}$ and generalize it 
into an approximation scheme for the systematic block diagonalization of $\bar{Q}$ into multiple (more 
than two) subspaces separated by spectral gaps. The generalized formalism is found to contain 
additional terms that do not appear in the standard scheme for bipartioning. The procedure is closely related to the 
dressed-atom approach described in reference \cite{AvanEtAl76} and has previously also been developed for the concrete 
example of a driven two-level system \cite{HausingerGrifoni10}.

Consider a Floquet system with quasienergy operator
\be
\bar{Q}=\bar{Q}_0+\lambda\bar{V}, \qquad \lambda = 1,
\ee
split into an uperturbed part $\bar{Q}_0$ and a perturbation $\lambda\bar{V}$. The dimensionless parameter
$\lambda$ shall eventually be set to one, and has been introduced to keep track of the order in which the 
perturbation appears.  The eigenstates of the unperturbed quasienergy operator $|\alpha m\rra$, the unperturbed Floquet 
modes, and their eigenvalues $\varepsilon^{(0)}_{\alpha m}$, the unperturbed quasienergies, are known and fulfill
\be\label{eq:Qeigen}
\bar{Q}_0|\alpha m\rra=\varepsilon^{(0)}_{\alpha m}|\alpha m\rra.
\ee
The index $m$ separates the eigenstates into multiple subsets. The states within each subset $m$ are 
labeled by the index $\alpha$ and span the unperturbed subspace $\mathcal{F}^{(0)}_m$ related to $m$. 
The quasienergies of two subsets $m$ and $m'$ shall be separated by a quasienergy gap that is large 
compared to the matrix elements of the perturbation $\bar{V}$.
When the perturbation is switched on smoothly, without closing the spectral gaps, the unperturbed 
subspaces $\mathcal{F}_m^{(0)}$ will be transformed adiabatically to the perturbed subspaces
$\mathcal{F}_m$, corresponding to a diagonal block of the perturbed problem. 
These subspaces $\mathcal{F}_m$ will be spanned by new basis states $|\alpha m\rra_B$ that deviate 
from the unperturbed states $\{|\alpha m\rra\}$ by small perturbative admixtures of states from other 
unperturbed subspaces. The states $|\alpha m\rra_B$ are decoupled from states that are not in~$\mathcal{F}_m$, 
\be
{}_B\lla\alpha'm'|\bar{Q}|\alpha m\rra_B = 0 \quad\text{for}\quad m'\ne m,
\ee
but generally they are no eigenstates of the perturbed quasienergy operator $\bar{Q}$, so that
${}_B\lla\alpha'm|\bar{Q}|\alpha m\rra_B$ can be finite also for $\alpha'\ne\alpha$. 
The task to be accomplished by degenerate perturbation theory is to find systematic expansions for 
both the perturbed basis states $|\alpha m\rra_B$, i.e.\ for the unitary operator $\bar{U}$ that 
relates them to the unperturbed basis states via 
\be\label{eq:Ualpham}
|\alpha m\rra_B=\bar{U}|\alpha m\rra,
\ee 
and the matrix elements ${}_B\lla \alpha' m|\bar{Q}|\alpha m\rra_B$ describing the physics within the
subspace~$\mathcal{F}_m$.

The projectors $\bar{P}_m$ into the unperturbed subspaces $\mathcal{F}^{(0)}_m$ are defined by 
\be
\bar{P}_m=\sum_{\alpha}|\alpha m\rra\lla \alpha m| 
\ee
and obey 
\be
\sum_m\bar{P}_m = \bar{1},
\ee
with the unity operator $\bar{1}$. They can be used to decompose any operator $\bar{A}$ like
\be
\bar{A}=\bar{A}_D +\bar{A}_X
\ee
into a block-diagonal part 
\be
\bar{A}_D = \sum_m \bar{P}_m\bar{A}\bar{P}_m
\ee
and a block-off-diagonal part
\be
\bar{A}_X =\sum_m \sum_{\substack{m'\\\ne m}}\bar{P}_{m'}\bar{A}\bar{P}_m.
\ee
The product of two block-diagonal operators is again block-diagonal,
\be
\bar{A}_D\bar{B}_D   =  ( \bar{A}_D\bar{B}_D )_D 
\qquad\text{i.e.}\qquad ( \bar{A}_D\bar{B}_D )_X=0,
\ee
and the product of a block-diagonal and a block-off-diagonal operator is block-off-diagonal,
\bes
\bar{A}_D\bar{B}_X &=& (\bar{A}_D\bar{B}_X)_X  
    \qquad\text{i.e.}\qquad ( \bar{A}_D\bar{B}_X )_D=0 ,
\\
\bar{A}_X\bar{B}_D &=& (\bar{A}_X\bar{B}_D)_X  
    \qquad\text{i.e.}\qquad ( \bar{A}_X\bar{B}_D )_D=0 .
\ees
If, like in the standard form of degenerate perturbation theory, the state space is bipartitioned only, one 
also finds $\bar{A}_X\bar{B}_X=(\bar{A}_X\bar{B}_X)_D$, but this relation does not hold if the state 
space is partioned into more than two subspaces. Instead, for multipartioning one generally has 
\be
\bar{A}_X\bar{B}_X   =  ( \bar{A}_X\bar{B}_X )_D + ( \bar{A}_X\bar{B}_X )_X.
\ee
The fact that the second term on the right hand side is finite will give rise to additional terms in 
the perturbation expansion that do not appear in the standard formalism.

We wish to block diagonalize the full unperturbed quasienergy operator $\bar{Q}$ by means of a unitary 
operator $\Uo$, such that 
\be\label{eq:QBD}
    \bar{U}^\dag\bar{Q}\bar{U}= \bar{Q}_0+\bar{W},
\ee
with block-diagonal operator
\be
\bar{W}=\bar{W}_D \qquad\text{i.e.}\qquad \bar{W}_X=0.
\ee
It is convenient to split Eq.~(\ref{eq:QBD}) into its block-diagonal part
\be\label{eq:QBDD}
[\bar{U}^\dag(\bar{Q}_0+\bar{V}_D+\bar{V}_X)\bar{U}]_D 
                    = \bar{Q}_0+\bar{W},
\ee
and its block-off-diagonal part
\be
\label{eq:QBDX}
[\bar{U}^\dag(\bar{Q}_0+\bar{V}_D+\bar{V}_X)\bar{U}]_X = 0.
\ee
The unitary operator $\bar{U}$ defines the new basis states via Eq.~(\ref{eq:Ualpham}). It can be expressed in terms of an anti-hermitian operator $\bar{G}$,
\be\label{eq:UG}
\bar{U}=\exp({\bar{G}}),
\qquad \bar{G}=-\bar{G}^\dag.
\ee 
However, $\bar{U}$ is not determined uniquely, unless one requires as the additional condition 
\be\label{eq:vV}
\bar{G}=\bar{G}_X \qquad\text{i.e.}\qquad \bar{G}_D=0
\ee
that keeps the mixing of states within each subpace $\mathcal{F}^{(0)}_m$ small. This ansatz defines 
the canonical van Vleck form of degenerate perturbation theory. 

We can expand $\bar{W}$ and $\bar{U}$ in powers $n$ of the perturbation $\bar{V}$,
\be
\bar{W} = \sum_{n=0}^\infty  \lambda^n \bar{W}^{(n)},
\ee
with
\be
\bar{W}^{(n)}=\bar{W}^{(n)}_D\qquad\text{i.e.}\qquad \bar{W}^{(n)}_X=0
\ee
and
\be
\bar{U} =  \sum_{n=0}^\infty \lambda^n  \bar{U}^{(n)}.
\ee
The terms $\Uo^{(n)}$ can be related to the terms in the perturbative expansion of $\bar{G}$,
\be
\bar{G} = \sum_{n=1}^\infty  \lambda^n \bar{G}^{(n)}
\ee
with
\be
\bar{G}^{(n)}=\bar{G}^{(n)}_X \qquad\text{i.e.}\qquad \bar{G}^{(n)}_D=0
\ee
and
\be
\bar{G}^{(n)}=-\big[\bar{G}^{(n)}\big]^\dag.
\ee
One has 
\bes
\bar{U}^{(0)} &=& 1,
\\
\bar{U}^{(1)} &=& \bar{G}^{(1)},
\\
\bar{U}^{(2)} &=& \bar{G}^{(2)}+\frac{1}{2}\big[\bar{G}^{(1)}\big]^2,
\\
\bar{U}^{(3)} &=& \bar{G}^{(3)}
            +\frac{1}{2}\big[\bar{G}^{(1)}\bar{G}^{(2)}+\bar{G}^{(2)}\bar{G}^{(1)}\big]
            +\frac{1}{6}\big[\bar{G}^{(1)}\big]^3,
\ees
etc.. Plugging these expressions into Eqs.~(\ref{eq:QBDD}) and (\ref{eq:QBDX}), we can iteratively 
determine $\bar{W}$ and $\bar{G}$ order by order. 

In zeroth order we find from Eq.~(\ref{eq:QBDD}) that
\be
\bar{W}^{(0)}=0,
\ee
while Eq.~(\ref{eq:QBDX}) reduces to $0=0$. For the next orders we obtain
\bes
\big[\bar{G}^{(1)},\bar{Q}_0\big] &=& \bar{V}_X
\\\label{eq:QG2}
{\big[\bar{G}^{(2)},\bar{Q}_0\big]} &=&     
    \big[\bar{V}_D,\bar{G}^{(1)}\big]   + \frac{1}{2}\big[\bar{V}_X,\bar{G}^{(1)}\big]_X
\\\label{eq:QG3}
{\big[\bar{G}^{(3)},\bar{Q}_0\big]} &=&     
    \big[\bar{V}_D,\bar{G}^{(2)}\big]
    +\frac{1}{3}\big[\big[\bar{V}_X,\bar{G}^{(1)}\big],\bar{G}^{(1)}\big]_X
\nonumber\\&&
    +\,\frac{1}{2}\big[\bar{V}_X,\bar{G}^{(2)}\big]_X
    -\frac{1}{4}\big[\big[\bar{V}_X,\bar{G}^{(1)}\big]_X,\bar{G}^{(1)}\big]_X
\ees
etc.\ from Eq.~(\ref{eq:QBDD}). And with that, Eq.~(\ref{eq:QBDX}) gives
\bes
\bar{W}^{(1)}&=&\bar{V}_D,
\\
\bar{W}^{(2)}&=&\frac{1}{2}\big[\bar{V}_X,\bar{G}^{(1)}\big]_D,
\\\label{eq:W3}
\bar{W}^{(3)}&=&\frac{1}{2}\big[\bar{V}_X,\bar{G}^{(2)}\big]_D 
            + \frac{1}{12}\big[\big[\bar{V}_X,\bar{G}^{(1)}\big],\bar{G}^{(1)}\big]_D,
\ees
where, in order to obtain the expression for $\bar{W}^{(3)}$, we have used that
$[[\bar{V}_X,\bar{G}^{(1)}]_X,\bar{G}^{(1)}]_D=[[\bar{V}_X,\bar{G}^{(1)}],\bar{G}^{(1)}]_D$, since
$[[\bar{V}_X,\bar{G}^{(1)}]_D,\bar{G}^{(1)}]_D=0$. For the terms $[\bar{G}^{(n)},\bar{Q}_0]$ the first 
deviation from the standard bipartioning perturbation theory appears in second order and is 
given by the second term on the right-hand side of (Eq.~\ref{eq:QG2}). Also the terms in the second line 
of Eq.~(\ref{eq:QG3}) are new. In the expansion of $\bar{W}$ the first deviation occurs in the third 
order $\bar{W}^{(3)}$ and is given by the last term of Eq.~(\ref{eq:W3}).

From the commutator $[\bar{G}^{(n)},\bar{Q}_0]$ we can construct all matrix elements of $\bar{G}$ order 
by order. Since $\bar{G}_D=0$, we know that 
\be\label{eq:ahm}
\lla \alpha' m|\bar{G}^{(n)}|\alpha m \rra =  0.
\ee
The non-vanishing matrix elements of $\bar{G}^{(n)}$ couple states $|\alpha m\rra$ and $|\alpha'm'\rra$ 
with $m'\ne m$ and obey 
\be
\lla \alpha'm'|\bar{G}^{(n)}|\alpha m\rra=-\lla \alpha m|\bar{G}^{(n)}|\alpha' m'\rra^* ,
\ee
following from $\bar{G}^{(n)}$ being anti-hermitian.  
By using that
\be
\lla\alpha'm'|\big[\bar{G}^{(n)},\bar{Q}_0\big]|\alpha m\rra = 
    (\varepsilon^{(0)}_{\alpha m}-\varepsilon^{(0)}_{\alpha' m'})
    \lla\alpha'm'|\bar{G}^{(n)}|\alpha m\rra,
\ee
one can compute the matrix elements with $m'\ne m$ of the leading terms:
\bes\label{eq:G1}
\lla \alpha'm'|\bar{G}^{(1)}|\alpha m\rra &=& 
        - \frac{\lla \alpha'm'|\bar{V}_X|\alpha m\rra}
            {\varepsilon^{(0)}_{\alpha' m'}-\varepsilon^{(0)}_{\alpha m}},
\\\label{eq:G2}
\lla \alpha'm'|\bar{G}^{(2)}|\alpha m\rra &=& 
\sum_{\alpha''} \frac{\lla \alpha'm'|\bar{V}_D|\alpha''m'\rra\lla\alpha''m'|\bar{V}_X|\alpha m\rra}
            {\big(\varepsilon^{(0)}_{\alpha' m'}-\varepsilon^{(0)}_{\alpha m}\big)
            \big(\varepsilon^{(0)}_{\alpha'' m'}-\varepsilon^{(0)}_{\alpha m}\big)}
\nonumber\\&& -\,
\sum_{\alpha''} \frac{\lla \alpha'm'|\bar{V}_X|\alpha''m\rra\lla\alpha''m|\bar{V}_D|\alpha m\rra}
            {\big(\varepsilon^{(0)}_{\alpha' m'}-\varepsilon^{(0)}_{\alpha m}\big)
            \big(\varepsilon^{(0)}_{\alpha' m'}-\varepsilon^{(0)}_{\alpha'' m}\big)}
\nonumber\\&& +\, 
        \sum_{\alpha''}\sum_{\substack{m''\\\ne m,m'}} 
        \frac{\lla \alpha'm'|\bar{V}_X|\alpha''m''\rra\lla\alpha''m''|\bar{V}_X|\alpha m\rra}
        {\varepsilon^{(0)}_{\alpha' m'}-\varepsilon^{(0)}_{\alpha m}}
\nonumber\\&&
        \times\frac{1}{2}\bigg[
        \frac{1}{\varepsilon^{(0)}_{\alpha'' m''}-\varepsilon^{(0)}_{\alpha' m'}}
        +\frac{1}{\varepsilon^{(0)}_{\alpha'' m''}-\varepsilon^{(0)}_{\alpha m}}\bigg].
\ees
In Eq.~(\ref{eq:G2}) one can explicitly see that the last term would not appear for a bipartition of 
the state space, since $m''$ must be different both from $m$ and $m'$.

When approximating the unitary operator $\bar{U}$ up to a finite order $n$ we have two possibilities, 
either
\be
\bar{U}\simeq\exp\Big(\bar{G}^{(1)}+\bar{G}^{(2)}+\cdots+\bar{G}^{(n)}\Big)\equiv\bar{U}^{[n]} 
\ee
or
\be
\bar{U}\simeq 1 + \bar{U}^{(1)}+\bar{U}^{(2)}+\cdots+\bar{U}^{(n)} .
\ee
Both approximations coincide up to order $n$ in the perturbation. However, the first approximation has 
the advantage that it preserves the unitarity, $\bar{U}^{[n]}$ is a unitary matrix also for finite $n$. In turn, the 
second approximation does not preserve unitarity for finite order. However, sometimes unitarity is not relevant and it is 
convenient to employ the second approximation to compute corrections to the perturbed basis states~(\ref{eq:Ualpham}) of 
the form 
\be
|\alpha m\rra_B=  \sum_{n=0}^\infty \lambda^n|\alpha m\rra_B^{(n)},
\qquad |\alpha m\rra_B^{(n)}= \bar{U}^{(n)}|\alpha m\rra.
\ee
Therefore, let us evaluate also the matrix elements of the leading terms $\bar{U}^{(n)}$:
\bes\label{eq:U0}
\lla \alpha'm'|\bar{U}^{(0)}|\alpha m\rra &=& \lla \alpha'm'|\alpha m\rra 
                = \delta_{\alpha'\alpha}\delta_{m'm}
\\\label{eq:U1}
\lla \alpha'm'|\bar{U}^{(1)}|\alpha m\rra &=& 
        - \frac{\lla \alpha'm'|\bar{V}_X|\alpha m\rra}
            {\varepsilon^{(0)}_{\alpha' m'}-\varepsilon^{(0)}_{\alpha m}},          
\\\label{eq:U3}
\lla \alpha'm'|\bar{U}^{(2)}|\alpha m\rra &=& 
\sum_{\alpha''} \frac{\lla \alpha'm'|\bar{V}_D|\alpha''m'\rra\lla\alpha''m'|\bar{V}_X|\alpha m\rra}
            {\big(\varepsilon^{(0)}_{\alpha' m'}-\varepsilon^{(0)}_{\alpha m}\big)
            \big(\varepsilon^{(0)}_{\alpha'' m'}-\varepsilon^{(0)}_{\alpha m}\big)}
\nonumber\\&& -\,
\sum_{\alpha''} \frac{\lla \alpha'm'|\bar{V}_X|\alpha''m\rra\lla\alpha''m|\bar{V}_D|\alpha m\rra}
            {\big(\varepsilon^{(0)}_{\alpha' m'}-\varepsilon^{(0)}_{\alpha m}\big)
            \big(\varepsilon^{(0)}_{\alpha' m'}-\varepsilon^{(0)}_{\alpha'' m}\big)}
\nonumber\\&& +\, 
        \sum_{\alpha''}\sum_{\substack{m''\\\ne m,m'}} 
        \frac{\lla \alpha'm'|\bar{V}_X|\alpha''m''\rra\lla\alpha''m''|\bar{V}_X|\alpha m\rra}
        {\big(\varepsilon^{(0)}_{\alpha' m'}-\varepsilon^{(0)}_{\alpha m}\big)
        \big(\varepsilon^{(0)}_{\alpha' m'}-\varepsilon^{(0)}_{\alpha'' m''}\big)} .
\nonumber\\&&
\ees

It is left to determine the matrix elements of the diagonal blocks of the quasienergy operator that 
describe the physics within the subspace $\mathcal{F}_m$, namely
\bes\label{eq:Qeff}
Q_{m,\alpha'\alpha}&=& {}_B\lla \alpha'm|\bar{Q}|\alpha m\rra_B 
                = \lla \alpha' m|\bar{U}^\dag\bar{Q}\bar{U}|\alpha m\rra 
\nonumber\\&=&       \lla \alpha' m|(\bar{Q}_0+\bar{W})|\alpha m\rra.
\ees
Expanding them in powers of the perturbation,
\be
Q_{m,\alpha'\alpha} = \sum_{n=0}^\infty  \lambda^n Q_{m,\alpha'\alpha}^{(n)},
\ee
we find the leading orders to be given by
\bes\label{eq:Qeff0}
Q_{m,\alpha'\alpha}^{(0)} &=& \lla \alpha'm|\bar{Q}_0|\alpha\rra 
        = \delta_{\alpha',\alpha}\varepsilon_{\alpha m}^{(0)},
\\\label{eq:Qeff1}
Q_{m,\alpha'\alpha}^{(1)} &=& \lla \alpha'm|\bar{V}_D|\alpha m\rra,
\\\label{eq:Qeff2}
Q_{m,\alpha'\alpha}^{(2)}&=&\sum_{\alpha''} \sum_{\substack{m'\\\ne m}} \lla \alpha'm|\bar{V}_X|\alpha''m'\rra\lla \alpha''m''|\bar{V}_X|\alpha m\rra
\nonumber\\&&
        \times\frac{1}{2}\bigg[
        \frac{1}{\varepsilon^{(0)}_{\alpha' m}-\varepsilon^{(0)}_{\alpha'' m'}}
        +\frac{1}{\varepsilon^{(0)}_{\alpha m}-\varepsilon^{(0)}_{\alpha'' m'}}\bigg],
\\\label{eq:Qeff3}
Q_{m,\alpha'\alpha}^{(3)} &=& 
\frac{1}{2}\sum_{\alpha''\alpha'''}\sum_{\substack{m'\\\ne m}}
        \Bigg\{
\nonumber\\&&
    \frac{\lla \alpha'm|\bar{V}_X|\alpha''m'\rra\lla \alpha''m'|\bar{V}_D|\alpha'''m'\rra
    \lla \alpha'''m'|\bar{V}_X|\alpha m\rra}
    {\big(\varepsilon^{(0)}_{\alpha'' m'}-\varepsilon^{(0)}_{\alpha m}\big)
        \big(\varepsilon^{(0)}_{\alpha''' m'}-\varepsilon^{(0)}_{\alpha m}\big)}
\nonumber\\&&+\,
    \frac{\lla \alpha'm|\bar{V}_X|\alpha''m'\rra\lla \alpha''m'|\bar{V}_D|\alpha'''m'\rra
    \lla \alpha'''m'|\bar{V}_X|\alpha m\rra}
    {\big(\varepsilon^{(0)}_{\alpha'' m'}-\varepsilon^{(0)}_{\alpha' m}\big)
        \big(\varepsilon^{(0)}_{\alpha''' m'}-\varepsilon^{(0)}_{\alpha' m}\big)}
\nonumber\\&&-\,
        \frac{\lla \alpha'm|\bar{V}_X|\alpha''m'\rra\lla \alpha''m'|\bar{V}_X|\alpha'''m\rra
        \lla \alpha'''m|\bar{V}_D|\alpha m\rra }
        {\big(\varepsilon^{(0)}_{\alpha'' m'}-\varepsilon^{(0)}_{\alpha''' m}\big)
        \big(\varepsilon^{(0)}_{\alpha'' m'}-\varepsilon^{(0)}_{\alpha m}\big)}
\nonumber\\&&-\,
        \frac{\lla \alpha'm|\bar{V}_D|\alpha''m\rra\lla \alpha''m|\bar{V}_X|\alpha'''m'\rra
        \lla \alpha'''m'|\bar{V}_X|\alpha m\rra }
        {\big(\varepsilon^{(0)}_{\alpha'' m}-\varepsilon^{(0)}_{\alpha''' m'}\big)
        \big(\varepsilon^{(0)}_{\alpha' m}-\varepsilon^{(0)}_{\alpha''' m'}\big)}
        \Bigg\}
\nonumber\\ &&+\,
\frac{1}{12}\sum_{\alpha''\alpha'''}\sum_{\substack{m'\\\ne m}}\;\sum_{\substack{m''\\\ne m,m'}}
        \Bigg\{
\nonumber\\&&
\lla \alpha'm|\bar{V}_X|\alpha''m'\rra\lla \alpha''m'|\bar{V}_X|\alpha'''m''\rra
        \lla \alpha'''m''|\bar{V}_X|\alpha m\rra 
\nonumber\\&&
\times\bigg[\frac{3}{\big(\varepsilon^{(0)}_{\alpha'' m'}-\varepsilon^{(0)}_{\alpha m}\big)
        \big(\varepsilon^{(0)}_{\alpha''' m''}-\varepsilon^{(0)}_{\alpha m}\big)} 
\nonumber\\&&-\,
\frac{3}{\big(\varepsilon^{(0)}_{\alpha'' m'}-\varepsilon^{(0)}_{\alpha m}\big)
        \big(\varepsilon^{(0)}_{\alpha'' m'}-\varepsilon^{(0)}_{\alpha''' m''}\big)}    
\nonumber\\&&-\,
\frac{3}{\big(\varepsilon^{(0)}_{\alpha' m}-\varepsilon^{(0)}_{\alpha''' m''}\big)
        \big(\varepsilon^{(0)}_{\alpha'' m'}-\varepsilon^{(0)}_{\alpha''' m''}\big)} 
\nonumber\\&&+\,
\frac{3}{\big(\varepsilon^{(0)}_{\alpha' m}-\varepsilon^{(0)}_{\alpha''' m''}\big)
        \big(\varepsilon^{(0)}_{\alpha' m}-\varepsilon^{(0)}_{\alpha'' m'}\big)}    
\nonumber\\&&+\,
\frac{1}{\big(\varepsilon^{(0)}_{\alpha'' m'}-\varepsilon^{(0)}_{\alpha''' m''}\big)
        \big(\varepsilon^{(0)}_{\alpha''' m''}-\varepsilon^{(0)}_{\alpha m}\big)}   
\nonumber\\&&+\,
\frac{1}{\big(\varepsilon^{(0)}_{\alpha' m}-\varepsilon^{(0)}_{\alpha'' m'}\big)
        \big(\varepsilon^{(0)}_{\alpha'' m'}-\varepsilon^{(0)}_{\alpha''' m''}\big)}            
\nonumber\\&&-\,
\frac{2}{\big(\varepsilon^{(0)}_{\alpha' m}-\varepsilon^{(0)}_{\alpha'' m'}\big)
        \big(\varepsilon^{(0)}_{\alpha''' m''}-\varepsilon^{(0)}_{\alpha m}\big)}           
\bigg]\Bigg\}.
\ees

\end{appendix}

\bibliography{mybib}

\end{document}